\documentclass[twocolumn]{aastex631}
\usepackage[mathlines]{lineno}
\usepackage{graphicx}   
\usepackage{lipsum} 
\usepackage{amsmath}
\usepackage{appendix}
\usepackage{float}

\usepackage{color}

\def\astrid{\texttt{ASTRID} }

\defcitealias{NG15-binary}{NG15-binary}
\defcitealias{NG15-evidence}{NG15-gwb}

\shorttitle{ASTRID Dual AGN}
\shortauthors{N.Chen et al.}
\graphicspath{{./}{figures/}}

\begin{document}

\title{Connecting current and future dual AGN searches to LISA and PTA gravitational wave detections}


\author{Nianyi Chen}
\affiliation{Max-Planck-Institut f\"ur Astrophysik, Karl-Schwarzschild-Str 1, D-85748 Garching, Germany}
\affiliation{School of Natural Sciences, Institute for Advanced Study, Princeton, NJ 08540, USA}

\author{Yihao Zhou}
\affiliation{McWilliams Center for Cosmology, Department of Physics, Carnegie Mellon University, Pittsburgh, PA 15213, USA}

\author{Ekaterine Dadiani}
\affiliation{McWilliams Center for Cosmology, Department of Physics, Carnegie Mellon University, Pittsburgh, PA 15213, USA}

\author{Tiziana Di Matteo}
\affiliation{McWilliams Center for Cosmology, Department of Physics, Carnegie Mellon University, Pittsburgh, PA 15213, USA}

\author{Cici Wang}
\affiliation{McWilliams Center for Cosmology, Department of Physics, Carnegie Mellon University, Pittsburgh, PA 15213, USA}

\author{Antonella Palmese}
\affiliation{McWilliams Center for Cosmology, Department of Physics, Carnegie Mellon University, Pittsburgh, PA 15213, USA}

\author{Yue Shen}
\affiliation{Department of Astronomy, University of Illinois at Urbana-Champaign, Urbana, IL 61801, USA}

\author{Junyao Li}
\affiliation{Department of Astronomy, University of Illinois at Urbana-Champaign, Urbana, IL 61801, USA}

\author{Adi Foord}
\affiliation{Department of Physics, University of Maryland Baltimore County, 1000 Hilltop Cir, Baltimore, MD 21250, USA}

\author{Simeon Bird}
\affiliation{Department of Physics \& Astronomy, University of California, Riverside, 900 University Ave., Riverside, CA 92521, USA}

\author{Yueying Ni}
\affiliation{Center for Astrophysics $\vert$ Harvard \& Smithsonian, Cambridge, MA 02138, USA}

\author{Yanhui Yang}
\affiliation{Department of Physics \& Astronomy, University of California, Riverside, 900 University Ave., Riverside, CA 92521, USA}

\author{Rupert Croft}
\affiliation{McWilliams Center for Cosmology, Department of Physics, Carnegie Mellon University, Pittsburgh, PA 15213, USA}

\correspondingauthor{Nianyi Chen}
\email{nianyi.chen7@gmail.com}

\begin{abstract}

Dual active galactic nuclei (DAGN) mark an observable stage of massive black hole (MBH) pairing in galaxy mergers and are precursors to the MBH binaries that generate low-frequency gravitational waves. Using the large-volume \texttt{ASTRID} cosmological simulation, we construct DAGN catalogs matched to current (COSMOS-Web, DESI) and forthcoming (AXIS, Roman) searches. With realistic selection functions applied, \texttt{ASTRID} reproduces observed dual fractions, separations, and host-galaxy properties across redshifts.  
We predict a substantial population of small-separation ($<5$ kpc) duals that current surveys fail to capture, indicating that the apparent paucity of sub-kpc systems in COSMOS-Web is driven primarily by selection effects rather than a physical deficit.
By following each simulated dual forward in time, we show that dual AGN are robust tracers of MBH mergers: $\sim$30--70\% coalesce within $\lesssim 1$ Gyr, and 20--60\% of these mergers produce gravitational-wave signals detectable by LISA. 
Duals accessible to AXIS and Roman are the progenitors of $\sim10$\% of low-redshift LISA events and $\sim30$\% of the PTA-band stochastic background. 
Massive green-valley galaxies with moderate-luminosity AGN, together with massive star-forming hosts containing bright quasars at $z>1$, emerge as the most likely environments for imminent MBH binaries. 
These results provide a unified cosmological framework linking dual AGN demographics, MBH binary formation, and gravitational-wave emission, and they identify concrete, high-priority targets for coordinated electromagnetic and GW searches in upcoming multi-messenger surveys.

\end{abstract}


\section{Introduction}
\label{section:introduction}

Hierarchical cosmological models predict the frequent occurrence of dual active galactic nuclei (AGN) as a natural outcome of galaxy mergers, when the central supermassive black holes (SMBHs) of the progenitor galaxies sink toward the center of the merger remnant and before forming a bound pair \citep[e.g.][]{Begelman1980, Volonteri2003ApJ...582..559V}. 
Dual AGN are therefore the parent population of interacting SMBHs, which will eventually coalesce and emit low-frequency gravitational waves (GWs). 
These GW signals are directly relevant for upcoming experiments, from the pulsar timing array (PTA) detection of the nanohertz gravitational-wave background \citep[GWB, e.g.][]{NG15_binary, EPTA2023A&A...678A..50E} to the future Laser Interferometer Space Antenna (LISA), which will detect SMBH binaries in the millihertz band \citep[e.g.][]{Amaro-Seoane2017arXiv170200786A}.

Despite their strong theoretical and astrophysical importance, the observational census of dual active galactic nuclei (AGN) remains limited and incomplete \citep[see, e.g., for recent comprehensive reviews][]{DeRosa2019, Pfeifle2024arXiv241112799P}. 
To date, only a few dozen dual AGN candidates have been robustly confirmed at kiloparsec-scale separations ($\lesssim 10,{\rm kpc}$), typically through targeted X-ray imaging or detailed spectroscopic follow-up observations \citep[e.g.][]{Silverman2020, Foord2021ApJ...907...71F, Mannucci2022NatAs...6.1185M, Shen2022, Li2024, Pfeifle2024arXiv241112799P, Perna2025A&A...696A..59P}. 
As a consequence, current samples remain too small to enable a statistically robust characterization of the fundamental properties of dual AGN systems, including their separation and mass distributions, accretion states, luminosity ratios, and host-galaxy properties. 
This observational limitation hinders our ability to place strong empirical constraints on theoretical models of massive black hole pairing, dynamical evolution, and merger timescales.

This situation is expected to improve substantially in the near future with the advent of several new and ongoing observational programs. 
In particular, the Advanced X-ray Imaging Satellite (AXIS) and the Nancy Grace Roman Space Telescope wide-field infrared survey (Roman) will dramatically expand the discovery space for dual AGN through their combination of high sensitivity, angular resolution, and large fields of view, enabling both efficient identification and statistical characterization of dual AGN populations across cosmic time. 
In parallel, the ongoing Dark Energy Spectroscopic Instrument (DESI) survey is obtaining high-quality spectra for nearly $10^5$ galaxies and quasars, providing a powerful dataset for the systematic identification of AGN pairs via spectroscopic diagnostics \citep[e.g.][]{Dadiani2025}. 
Together, these forthcoming and ongoing datasets bring significant advances in our understanding of the demographics and environments of dual AGN, and will play a key role in strengthening the synergy between electromagnetic observations of massive black hole binaries and gravitational-wave searches.

On the theoretical side, cosmological hydrodynamical simulations have recently advanced to the point of resolving both the galaxy population and the black hole dynamics required to study dual AGN formation and evolution. 
Recent efforts have used state-of-the-art simulations to construct mock catalogs of dual AGN, investigate their observability, and link them to SMBH binary populations \citep[e.g.][]{Volonteri2022, Chen2023, LiKunyang2023ApJ...959....3L, Saeedzadeh2024ApJ...975..265S, Puerto-Sanchez2025MNRAS.536.3016P}.
Such forward-modeling efforts are crucial for interpreting survey data, connecting electromagnetic observations to GW source populations, and predicting the multi-messenger signal of SMBH pairs.
However, most of these studies use generic dual AGN selection criteria (e.g., a spatial separation below $\Delta r < 30\,{\rm kpc}$ and AGN bolometric luminosity of $L_{\rm bol} > 10^{43}\,{\rm erg/s}$) for dual AGN selection from the simulation.
Such methods hinder direct comparison with observation catalogs or the interpretation of the comparisons, as the selection function has a large impact on the resulting dual AGN population and how they evolve into SMBH binaries.
It is crucial, given the diversity of ongoing and upcoming dual AGN searches, to create simulated dual AGN catalogs tailored to each observation sample, to better understand the SMBH pairs probed by each search strategy, and to bridge the gap between the DAGN simulation and observations.

In this work, we make use of the \texttt{ASTRID} simulation, a large-volume cosmological simulation with high spatial and mass resolution, to construct mock catalogs of dual AGN matched to the selection functions of DAGN searches using COSMOS-Web and DESI data.
We also generate predictive dual AGN catalogs for upcoming observation samples from AXIS and Roman.
We investigate the statistical properties of these dual AGN samples, their relation to the underlying SMBH binary population, and their predicted contribution to the PTA-band GWB and LISA-detectable GW sources. 
By bridging large-scale galaxy surveys and gravitational-wave astronomy, we aim to provide a coherent framework for interpreting dual AGN observations as precursors of GW events.

The paper is organized as follows.
In Section \ref{sec:simulation}, we briefly introduce the \texttt{ASTRID} simulation, with a focus on the BH modeling.
Section \ref{sec:catalog} describes the selection criteria for generating mock catalogs tailored to each observation sample.
In Section \ref{sec:pop}, we show the properties of dual AGN across redshifts and in different mock samples, with direct comparisons to the observed dual candidates.
Finally, section \ref{sec:merger} investigates the connection between dual AGN population and SMBH binaries in the PTA and LISA band, and the host galaxies of dual AGN mergers.

\section{Simulation}
\label{sec:simulation}
We generate dual AGN mock catalogs using the AGN population from the \texttt{ASTRID} simulation. 
\texttt{ASTRID} contains models for galaxy formation, including gas cooling, star formation, metal return, BH seeding, merging, and accretion as well as supernova and AGN feedback (including both thermal and kinetic modes).
\texttt{ASTRID} is fully described in the introductory papers \citep[e.g.][]{Bird-Astrid, Ni-Astrid, Chen2022, Chen2023, DiMatteo2023, Ni2024, Zhou2025_astridz0}. Here, we briefly introduce the basic parameters and BH modeling.
\texttt{ASTRID} runs in a uniquely large $250\, {\rm cMpc}/h$ box with $2\times 5500^3$ dark matter+gas particles, with a 
dark matter resolution element of mass $M_{\rm DM} = 9.94 \times 10^6\,M_\odot$. 
The gravitational softening length is $\epsilon_{\rm g} = 1.5 \, {\rm ckpc}/h$. 

The large volume of \texttt{ASTRID} enables us to investigate rare pairs of bright AGN with $L_{\rm bol} >10^{45}\,{\rm erg/s}$ \citep[e.g.][]{Shen2022, Chen2023}. 
In \texttt{ASTRID}, BHs are seeded in halos with $M_{\rm halo} > 5 \times 10^9 M_\odot/h$ and $M_{\rm *} > 2 \times 10^6 M_\odot/h$, with seed masses stochastically drawn from a power law between $3\times10^{4} \,M_\odot/h$ and $3\times10^{5} M_\odot/h$. 
When the accretion rate is less than $5\%$ of the Eddington ratio and $M_{\rm BH} \gtrsim 10^{8.5} M_\odot/h$, the BHs enter a kinetic feedback mode, similar to \citet{Weinberger2017}, but with parameters that make kinetic feedback moderately less aggressive \citep{Ni-Camels}. 

Uniquely among large-scale cosmological simulations, \texttt{ASTRID} includes a subgrid dynamical friction model \citep{Tremmel2015,Chen2021} from dark matter, stars, and gas, yielding physically consistent black hole trajectories and velocities.
This model, compared to the repositioning of MBHs, avoids spurious BH mergers and resolves dynamical friction to the numerical spatial resolution limit (gravitational softening length of $\epsilon_g = 1.5\,{\rm ckpc/h}$).
This is particularly crucial for robust modeling of kpc-scale dual AGN.
Two black holes merge when their separation is within two times the gravitational softening length $2\epsilon_g=3\,{\rm ckpc}/h$, and if their kinetic energy is dissipated by dynamical friction and they are gravitationally bound to the local potential.

\section{Dual AGN Catalogs}
\label{sec:catalog}
\begin{figure*}
\centering
\includegraphics[width=0.8\textwidth]{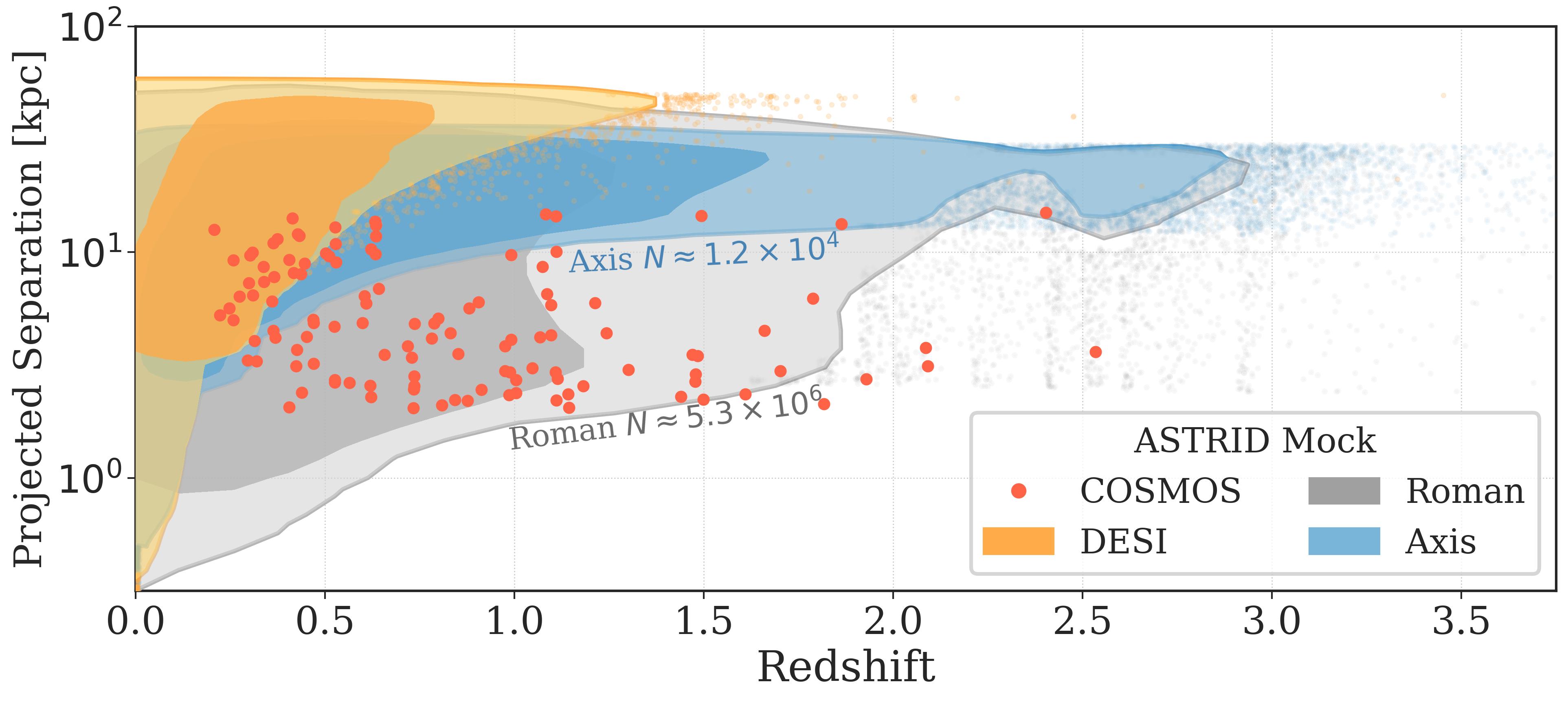}
  \caption{Redshift and projected separation of light-cone samples in each mock dual AGN catalog. We mock four observational samples: the COSMOS-Web sample \citep[\textit{red dots},][]{LiJunyao2025ApJ...986..101L} and the DESI sample \citep[\textit{orange},][]{Dadiani2025} are mocking existing dual AGN candidates from recent observations; the AXIS (\textit{blue}) and Roman (\textit{grey}) mocks target two future observations with the potential to discover a large population of dual AGN. We also label the (optimistic) estimation for the number of dual AGN candidates discoverable by AXIS and Roman.}
  \label{fig:dr_z_sel}
\end{figure*}

\begin{figure}
\centering
\includegraphics[width=0.48\textwidth]{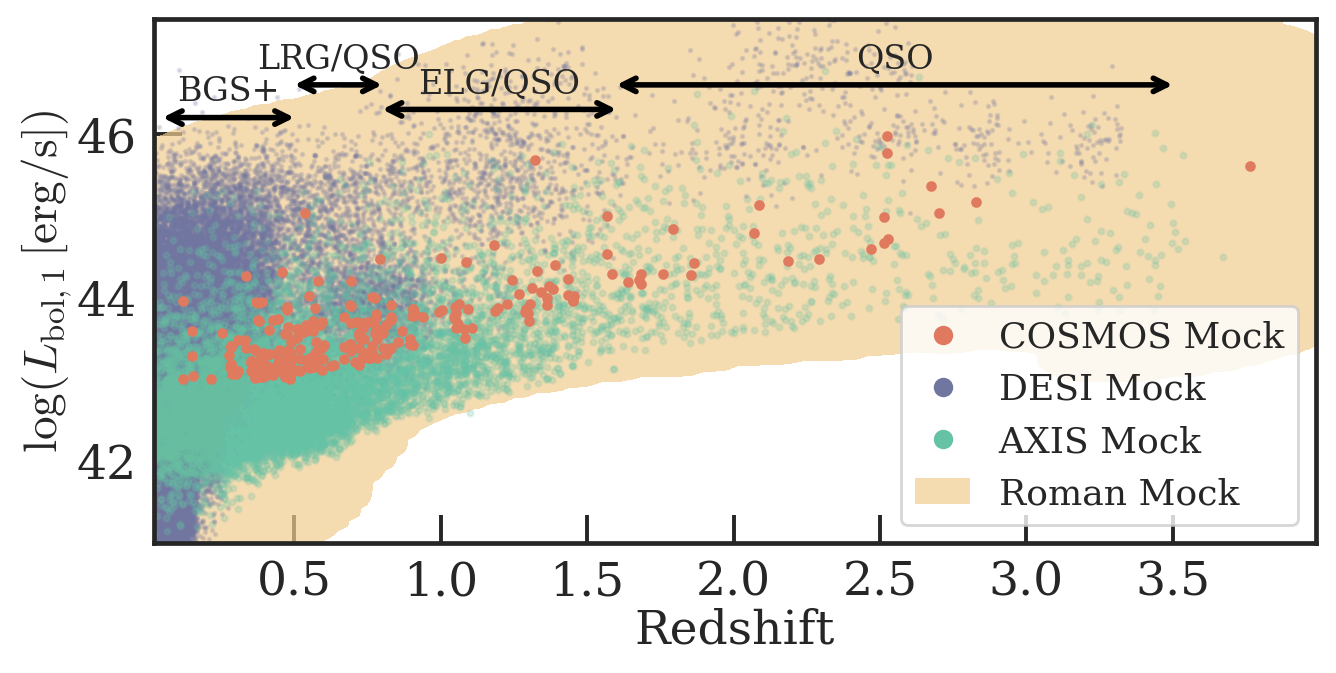}
  \caption{Redshift and bolometric luminosity of the primary AGN of the four mock samples. The black arrows and text label the DESI dual AGN categories in each redshift range.}
  \label{fig:lbol_z_sel}
\end{figure}

From \texttt{ASTRID}, we generate four dual AGN mock catalogs with different selection criteria (based on AGN luminosity, host galaxy properties, and separation) to match four observation samples, for a direct comparison with state-of-the-art observations.
The four targeted observations are existing dual AGN candidates from  COSMOS-Web \citep{LiJunyao2025ApJ...986..101L} and DESI \citep{Dadiani2025}, along with two potential dual AGN searches using AXIS and Roman.
The parameter space covered by these four mock dual AGN catalogs is illustrated in Figure \ref{fig:dr_z_sel} and Figure \ref{fig:lbol_z_sel}.

Figure \ref{fig:dr_z_sel} shows the redshift and projected separation distribution of each mock catalog.
We see that among the four searches, Roman has the highest spatial resolution to probe dual AGN below kpc separations.
We also estimate the total number of duals that AXIS and Roman will be able to find. 
We note that these are optimistic estimations, as we do not fully account for the AGN obscuration and host galaxy contamination.
Figure \ref{fig:lbol_z_sel} shows the primary AGN (the more luminous one in the dual) luminosity and redshift distribution.
In the following subsections, we present the detailed selection criteria for the four mock dual AGN catalogs.

\begin{figure*}
\centering
\includegraphics[width=1\textwidth]{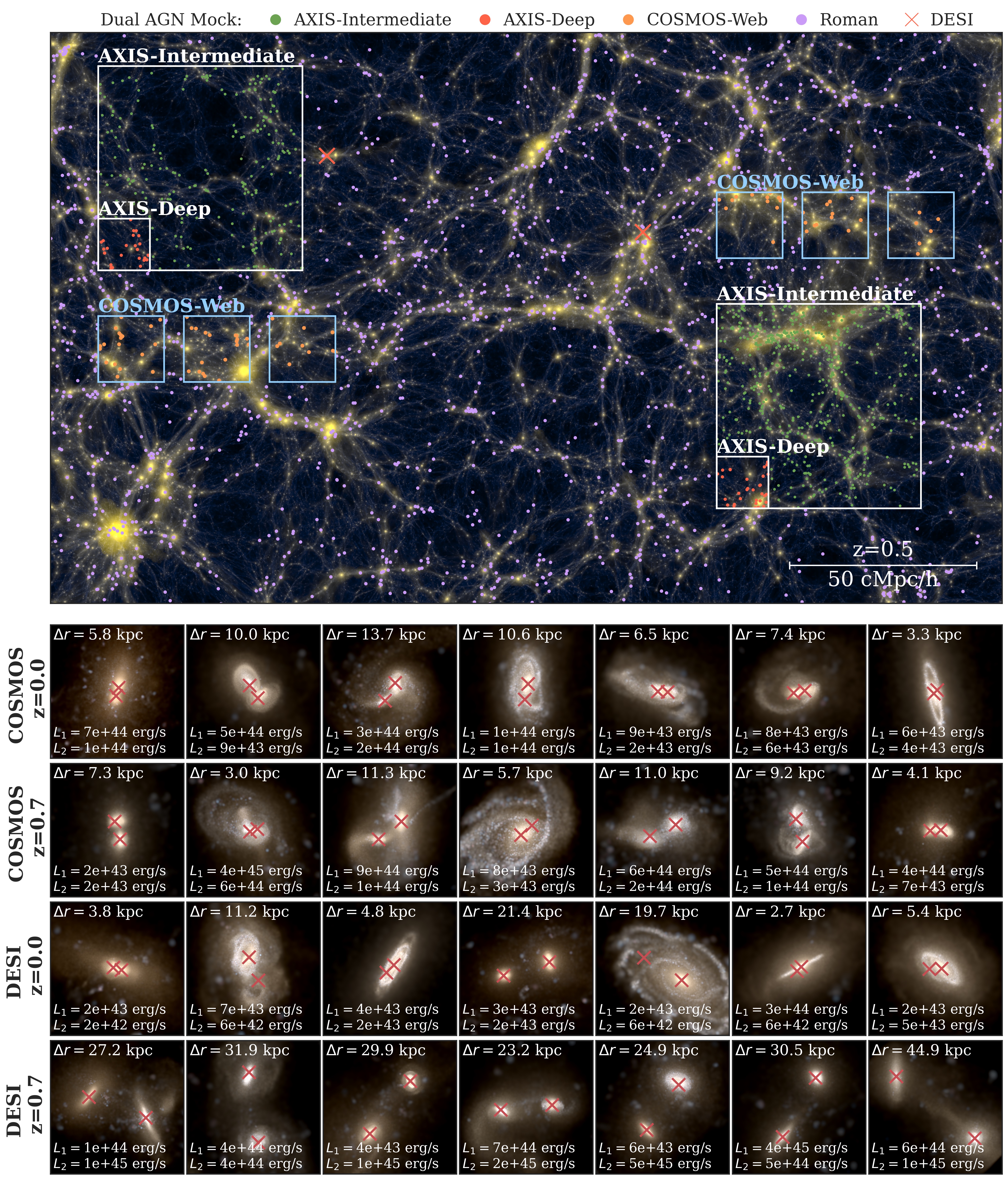}
  \caption{\textit{Upper:} $z=0.5$ gas density colored by temperature in a slice of $250\ {\rm Mpc}/h\times 150\ {\rm Mpc}/h$. Duals within a slab of $20\ {\rm Mpc/h}$ in thickness are marked. For COSMOS and AXIS, we only present the mock duals in the squares (COSMOS volume: 0.26 ${\rm deg}^2$; AXIS-Deep volume: 0.16 ${\rm deg}^{2}$;
  AXIS-Intermediate volume: 2.5 ${\rm deg}^{2}$;
  )
  \textit{Lower:} each panel is 60 ckpc$/h$ per side. The projected separation (in physical units) of each dual is labeled. }
  \label{fig:lss_slice}
\end{figure*}

\subsection{Dual AGN in the COSMOS-Web Field}
\label{sec:cosmos-mock}
Systematic searches for dual AGN across redshifts and down to kpc-scales have been challenging.
Recently, \cite{LiJunyao2025ApJ...986..101L} presented a statistical sample of dual AGN candidates from $z=0.5-5$ with separations as low as $2\,~{\rm kpc}$.
They searched for AGN companions around 571 moderate luminosity, X-ray-selected AGNs using deep HST ACS/F814W and multiband JWST NIRCam imaging from the COSMOS-Web survey, and identified 59 dual AGN candidates based on spatially resolved spectral energy distribution analyses.
This sample is very suitable for a statistical comparison with simulations, because of the relatively straightforward luminosity cut based on X-ray luminosity and the uniform separation selection.

It has been shown in \cite{LiJunyao2025ApJ...986..101L} that the dual/pair fraction from the \texttt{ASTRID} and \texttt{Horizon-AGN} cosmological simulations generally agrees with their observed dual/pair fraction, although there remains some discrepancy towards low redshifts.
A somewhat surprising finding of \cite{LiJunyao2025ApJ...986..101L} is a flat separation distribution among duals between $2-15\,{\rm kpc}$, and a lack of pairs below $2\,{\rm kpc}$ despite the instrument's ability to resolve them, whereas simulations predict that duals accumulate at $\Delta r<{10}\,{\rm kpc}$ \citep[e.g.][]{Volonteri2022, Chen2023}.
We aim to more carefully quantify the level of consistency between the COSMOS-Web dual candidates and the simulated sample by mocking both their AGN and galaxy properties and the sample size, to capture the role of cosmic variance and Poisson fluctuations.

We first generate a sample of single and dual AGN at $z=0- 4$ matching the luminosities and separations of their sample (\texttt{COSMOS-mock} hereafter).
We select single AGN at each redshift based on the minimum luminosity of the single AGN in the \cite{Li2024} sample (see their Figure 1).
Then, we search for a companion AGN with projected separations $2\,{\rm kpc} < r_{\rm 2D} < 15\,{\rm kpc}$ around the single AGN.
We require the companion's bolometric luminosity to be larger than $1/10$ of the main AGN luminosity, matching the observed sample.
In case of AGN tuples with $N>2$, we take the brightest two AGN as the dual.
The multiple AGN population constitutes $\sim 10\%$ among the dual AGN sample.

\subsection{Dual AGN in DESI DR1}
\label{sec:desi-mock}
Despite the wide luminosity range and high spatial resolution probed by the COSMOS-Web sample, the search area of $0.26\,{\rm deg}^2$ limits the total number of pairs in this field, so statistical uncertainties due to Poisson fluctuations remain large. 
Recently, a large sample of dual AGN candidates from the DESI Data Release 1 (DR1; \citealt{DESI_DR1}) AGN population (Juneau et al. in preparation), which also includes data from the DESI Early Data Release \citep{DESI:2023ytc}, could be identified \citep{Dadiani2025}. 
This sample has a much wider survey area of $9528\,{\rm degree}^2$ (or 23\% of the full sky) and probes fainter AGN at low redshifts \citep[e.g.][]{Pucha2025ApJ...982...10P}. 

DESI DR1 provides spectroscopic measurements for approximately 17 million galaxies and quasars across four DESI target classes \citep[Bright Galaxy Survey, or BGS, Luminous Red Galaxies, or LRG, Emission Line Galaxies, or ELG, QSO, and secondary programs;][]{2023AJ....165...50M}, making it one of the largest spectroscopic datasets currently available. Spectral classifications and redshifts are obtained with the Redrock pipeline (Bailey et al., in prep), while quasar identifications and redshifts are further improved using QuasarNet and a Mg II-based refinement procedure \citep{2023ApJ...944..107C, 2023AJ....165..124A}.
AGN diagnostics combine optical/UV emission-line measurements from FastSpecFit \citep{2023ascl.soft08005M} with mid-infrared WISE photometry processed with Tractor \citep{2016ascl.soft04008L}. 
AGN host-galaxy properties are estimated using spectral energy distribution fitting with the Code Investigating GALaxy Emission \citep{2019A&A...622A.103B, 2024A&A...691A.308S}. 
This comprehensive spectroscopic and multi-wavelength characterization enables a systematic and robust search for close AGN pairs.
Further description of the dual AGN population in DR1 is provided in \cite{Dadiani2025}.

We compensate the COSMOS-Web Mock with a mock dual AGN catalog targeting the dual AGN population in DESI DR1.
Because the DESI survey targets different tracers across redshifts, it is important to match the AGN population in different galaxy types when creating the DESI mock dual AGN sample.
We divide the sample into four redshift bins and match the major DESI AGN population in each bin.
At $z\geq 1.6$, the DESI AGN population is dominated by QSOs. 
To match the QSO tracer in DESI, we select the AGN in \texttt{ASTRID} which outshines their host galaxy in the UV band. 
To estimate the UV-band luminosity of AGNs, we follow \citet{Fontanot2012} and apply bolometric corrections to convert $L_{\rm bol}$ to rest-frame UV-band absolute magnitude $M_{\rm UV}$:
\begin{equation}
    M_{\rm UV} = -2.5\log_{10} \frac{L_{\rm Bol}}{6.83\times 10^{15}\ {\rm J/s}}-33.62.
\end{equation}
The UV luminosity of the host galaxy is calculated by constructing its spectral energy distribution. We model each star particle as a simple stellar population (SSP) with its birth time, metallicity, and mass extracted from the simulation. We use the \textsc{FSPS} stellar population synthesis code \citep{Conroy2009, Conroy2010} with the PARSEC isochrones \citep{Bressan2012_parsec}, MILES stellar library \citep{Sanchez-Blazquez2006_miles}, assuming a Chabrier initial mass function \citep{Chabrier2003}. 
The luminosity for an individual galaxy is the sum of the emission of all star particles in this galaxy.  

At $0.8\leq z<1.6$, we select the dual AGN from the ELG-like sample and QSO, where our ELG-like sample is selected with ${\rm SFR} > 1\,M_\odot/{\rm yr}$ and $10^{8.5}\,M_\odot <M_{\rm gal} < 10^{10.5}\,M_\odot$ \citep[e.g.][]{Yuan2025MNRAS.538.1216Y}.
At $0.5\leq z<0.8$, the DESI dual AGN population is dominated by pairs from LRG and QSO, and we use $M_{\rm gal} > 10^{11}$ $M_\odot$ to represent the LRG population \citep[e.g.][]{Yuan2022MNRAS.512.5793Y}.
Finally, at $z<0.5$, the DESI dual sample contains a mixture of BGS sources and secondary targets.
We directly match the single AGN luminosity distribution at these redshifts and select duals from this single AGN population.
The resulting bolometric luminosity of \texttt{DESI-Mock} duals is shown in dark blue dots in Figure \ref{fig:lbol_z_sel}.
We select duals with minimum projected separations of $2''$ and maximum separations of $50\,{\rm kpc}$ following \cite{Dadiani2025}.

\subsection{Dual AGN in Next-generation X-ray Observatory: AXIS}
Advanced X-ray Imaging Satellite (AXIS) is a proposed NASA Probe-class mission that will provide high angular resolution X-ray imaging and spectroscopy over a large field of view \citep{Reynolds2024SPIE13093E..28R}.
With sensitivity ten times greater than that of the Chandra X-ray Observatory and image resolution of $\sim 1''$, AXIS can probe dual AGN with $L_{\rm bol}\sim 10^{41}\,{\rm erg/s}$ at $z=0.1$ and $L_{\rm bol}\sim 10^{43}\,{\rm erg/s}$ at $z=3$, significantly fainter than most current dual AGN samples.
Moreover, optical selection techniques for AGNs are affected by optical extinction and contamination from star formation, which is especially problematic when observing highly obscured mergers. 
X-ray observations are less affected by these issues, making AXIS an ideal instrument for identifying dual AGN and confirming candidates identified in other wavelengths \citep[e.g.][]{Ricci2021MNRAS.506.5935R}.

We create a mock dual AGN catalog targeting the AXIS intermediate survey, with a total survey area of $\sim 2.5\,{\rm deg}^2$ and a flux limit of $F_{\rm 2-10 keV} = 2\times 10^{-16}\,{\rm erg/s/cm^2}$ \citep{Marchesi2020A&A...642A.184M, Cappelluti2024Univ...10..276C}.
We use the lower $L_x$ limit for the simulated AXIS AGN at each redshift \citep{Foord2024Univ...10..237F} and convert it to a minimum bolometric luminosity, assuming a bolometric correction of 10 for the 2-10 keV band.
We select dual AGN with projected separations $r(\theta=1'') < r_{\rm 2D} < 30\,{\rm kpc}$, which can be resolved by AXIS at $z>0.1$.

\subsection{Dual AGN in Future Infrared Survey: Nancy Grace Roman Space Telescope}
Our last mock dual AGN catalog targets the future Nancy Grace Roman Space Telescope wide-field infrared survey \citep{Spergel2015arXiv150303757S}.
The Roman telescope will conduct a High Latitude Survey (HLS) covering $\sim 2000\,{\rm deg}^2$ in multiple NIR bands down to a depth of $\sim 26.9$ AB mag \citep{Akeson2019arXiv190205569A}.
With its high spatial resolution of $\sim 0.1''$ and wide survey area, Roman has the potential to identify a large number of dual AGN candidates across cosmic time, down to kpc-scale separations.
We create a mock dual AGN catalog (\texttt{Roman-mock} hereafter) targeting this future survey.

We select dual AGN with projected separations of $0.1'' < r_{\rm 2D} < 30\,{\rm kpc}$, corresponding to the spatial resolution limit and a typical maximum separation for dual AGN.
We note that the minimum separation resolvable by Roman is smaller than the gravitational softening length of \texttt{ASTRID} at $z<1$, and so our results at these redshifts may be affected by resolution effects.
To estimate the minimum bolometric luminosity of AGN detectable by Roman at each redshift, we convert the survey depth of 26.9 AB mag in the $H$-band to a flux limit, and then to a bolometric luminosity assuming the AGN SED template from \cite{Lyu2017ApJ...835..257L}.
We select dual AGN with bolometric luminosities above this limit at each redshift.

\subsection{Light-cone Mock Catalogs}

When generating dual AGN catalogs from simulations and comparing with observations, most previous works \citep[e.g.][]{DeRosa2019, Volonteri2022, Chen2023} extract single and dual AGN from each simulation snapshot, and with the volume of the full simulation box. 
A drawback of this approach, especially when comparing with small field-of-view observations, is that they do not account for the cosmic variance and Poisson fluctuation and may overestimate the level of inconsistency with observed dual statistics.
Furthermore, the light-cone is necessary self-consistently link the simulation dual AGN with MBH merger rates and merger properties.
For this work, we generate mock light-cone dual catalogs for each observation, in addition to the snapshot-level dual catalogs.
Here we briefly describe our light-cone samples.

In each simulation snapshot with a cadence of $\Delta z=0.1$, we first generate a catalog of single AGN and dual AGN in the full simulation volume following the selection criteria outlined in previous sections.
Then, starting from $z_1=0$, we find $z_2$ such that the comoving distance between $z_1$ and $z_2$ equals the simulation box size.
We extract all single AGN and dual AGN in the corresponding snapshot (at redshift closest to $z_{mid}=(z_1 + z_2)/2$), inside a ``cone'' along the z-axis, with a randomly sampled center in the x-y plane, and an area equal to the angular area at the corresponding redshift.
We then set $z_1=z_2$, and proceed to the next redshift bin, until we get to $z=3$.
For COSMOS-Web and AXIS, the survey volume is completely contained in the simulation box size, and we use the approximate survey area ($0.2$ and $2\,{\rm deg}^2$ respectively) for the light-cone sampling.
For DESI and Roman, the survey area far exceeds the simulation box size, especially at high redshifts. 
Hence, we generate light-cones with angular area of $4\,{\rm deg}^2$ (the extent of our box at $z\sim 3$).

For each mock sample, we randomly sample $1000$ light-cones.
We use them to account for the statistical uncertainties due to cosmic variance and Poisson fluctuations in COSMOS-Web and AXIS.

\section{Properties of the Dual AGN Population}
\label{sec:pop}

\subsection{Dual AGN Fraction}

\begin{figure}
\centering
\includegraphics[width=0.44\textwidth]{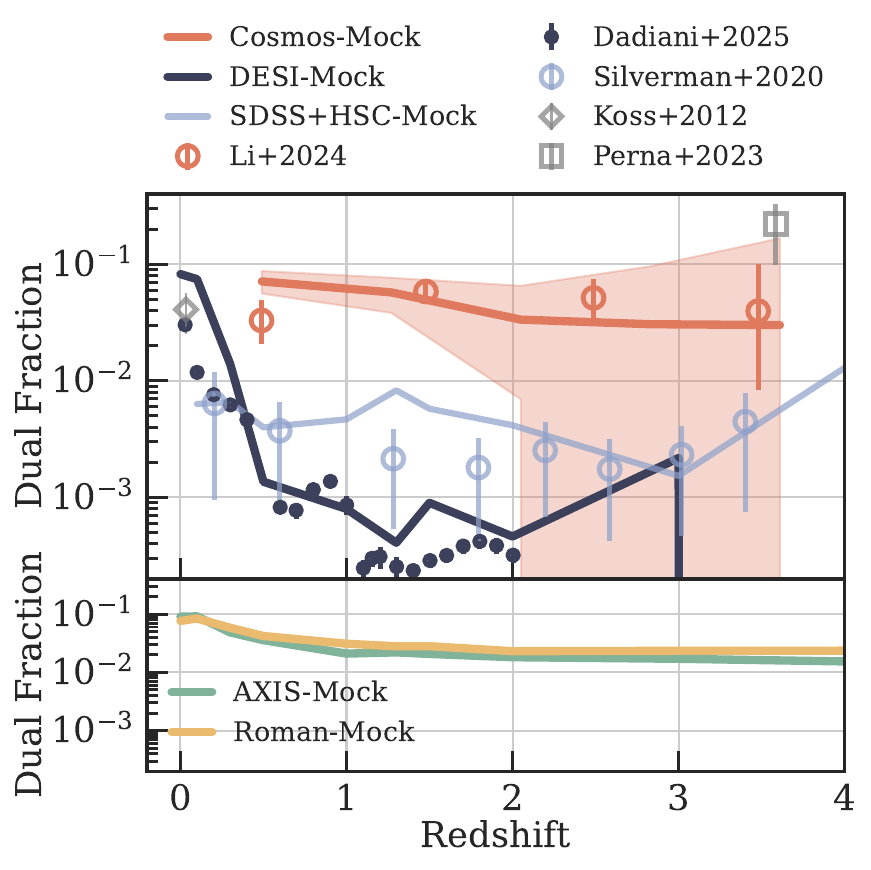}
  \caption{\textit{\textbf{Top:}} Dual AGN fraction as a function of redshift for dual AGN mock catalogs  created from \texttt{ASTRID}, targeting the COSMOS-Web sample \citep[][\textit{red}]{LiJunyao2025ApJ...986..101L}, SDSS+HSC sample \citep[][\textit{light purple}]{Silverman2020} and DESI AGN sample \citep[][\textit{dark blue}]{Dadiani2025}. For the COSMOS-Web mock we also show the $2\sigma$ uncertainty due to cosmic variance and Poisson fluctuation from 200 light-cone realizations (\textit{red shaded}). The dual AGN fraction in each observed sample is shown as circles in the corresponding colors. We also show the dual fraction quoted in \cite{Perna2025A&A...696A..59P} at $z=3.5$ and \cite{Koss2012} at $z\sim 0$. 
  Notably, the high dual AGN fraction measured by \cite{Perna2025A&A...696A..59P} is marginally outside the $2\sigma$ of the simulated sample distribution with similar luminosities. \textit{\textbf{Bottom:}} predicted dual AGN fraction for upcoming observations AXIS and Roman.
  }
  \label{fig:dual_frac}
\end{figure}

The fraction of dual AGN among the overall AGN population is a key observable that encodes information about the triggering of SMBH activity during galaxy mergers, as well as the dynamical evolution of MBH pairs.
A comparison of dual AGN fractions between different observations with a generic simulated sample can be misleading, as different observational surveys use different AGN samples and companion searching methods, and thus yield dual fractions spanning one order of magnitude.
For example, \cite{Silverman2020} targets duals among luminous quasars in SDSS and searches for double nuclei through HSC imaging and found a dual fraction of $\sim 0.06\%$ at $z=0.8-4.5$.
\cite{LiJunyao2025ApJ...986..101L} searched for a second nucleus around Chandra-selected AGN using JWST, and found a much higher dual fraction of $\sim 7\%$ at $z=0.75-2.5$.
This is due to the different luminosity thresholds, sensitivity to obscured duals, and spatial resolution probed by the two surveys, where the COSMOS-Web sample probes lower luminosity AGN and smaller separations, both of which can increase the probability of finding a dual AGN companion.
It is thus crucial, given the variation in the resulting dual fraction from different selection criteria, to match the simulated sample closely to each observation selection function before a meaningful comparison can be made.

The top panel of Figure \ref{fig:dual_frac} shows the dual AGN fraction from recent observations in the redshift range of $z=0-4.5$, \cite{Silverman2020} (SDSS+HSC sample), \cite{LiJunyao2025ApJ...986..101L} (COSMOS-Web sample), \cite{Dadiani2025} (DESI sample), \cite{Perna2025A&A...696A..59P} (COSMOS sample at $z\sim 3.5$), and \cite{Koss2012} (Swift/BAT sample at $z\sim 0$), compared with the dual AGN fractions from our mock catalogs tailored to each observational sample.
We define the dual AGN fraction as the number of dual AGN divided by the number of AGN that match the primary AGN selection criteria among duals.
In addition to the COSMOS-Web and DESI mock catalogs described in Section \ref{sec:catalog}, we also mock the dual quasar sample in \cite{Silverman2020} with $\log(L_{\rm bol,1}[\rm erg/s]) >45.3$, $L_{\rm bol,2} > L_{\rm bol,1}/10$, $5\,{\rm kpc} < \Delta r_{\rm 2D} < 30\,{\rm kpc}$, and two distinguishable galaxy bulges.
We also note that there is a recent independent measurement of dual fraction using SDSS quasars by \cite{Jiang2025arXiv250406415J}, who found a lower dual fraction than \cite{Silverman2020}, and pointed out that there could be contamination by stars in \cite{Silverman2020}.

There is generally good agreement between simulation predictions and observational constraints when similar selection criteria are applied.
We demonstrate using the same underlying physical model that the dual AGN fraction strongly depends on both the luminosity threshold and the spatial resolution limit of the survey.
The high AGN luminosity cut for both AGN in the SDSS+HSC sample requires a major merger between two quasar hosts, while the COSMOS-Web sample is sensitive to lower luminosity ratio pairs, which are more common.
The dependence on selection is also reflected in the DESI sample, where the high-z quasar-dominated pairs are rarer than the low-z AGN pairs in LRGs and dwarf galaxies.

We also note the surprisingly high dual AGN fraction of $\sim 20\%$ measured by \cite{Perna2025A&A...696A..59P} at $z\sim 3.5$.
This fraction may indicate a stronger enhancement of dual AGN triggering at high redshifts, or more clustering of AGN at these epochs.
However, given the small sample size of $3-5$ dual candidates among 16 AGN, interpreting this high dual fraction requires quantifying the statistical uncertainties due to cosmic variance and Poisson fluctuations.
For example, measurements using X-ray-selected dual AGN put an upper limit of $\sim 4\%$ to the dual AGN fraction at $z\sim 3$ \citep[e.g.][]{Sandoval2024ApJ...974..121S}. 
To this end, we create 200 light-cone realizations of the COSMOS-Web mock catalog, which has luminosity and spatial resolution limits similar to the \cite{Perna2025A&A...696A..59P} sample.
Each mock realization contains $\sim 10-30$ single AGN at $z>3$, from which we calculate the dual AGN fraction.
The distribution of the dual AGN fraction from these realizations is shown as the red shaded region in Figure \ref{fig:dual_frac}.
We find that the high dual AGN fraction measured by \cite{Perna2025A&A...696A..59P} lies at the $2\sigma$ upper bound of the simulated distribution, indicating that statistical fluctuations may play a significant role in producing the high dual fraction in this small sample.

In the bottom panel of Figure \ref{fig:dual_frac}, we show the predicted dual AGN fractions for future surveys with AXIS and Roman, based on our mock dual AGN catalogs targeting these surveys.
We find a flat dual fraction of $\sim 2-3\%$ for both surveys across $z=1-4$, and an increase towards $10\%$ at $z<1$, when both surveys become increasingly sensitive to lower luminosity AGN in dwarf galaxies.

\begin{figure*}
\centering
\includegraphics[width=0.98\textwidth]{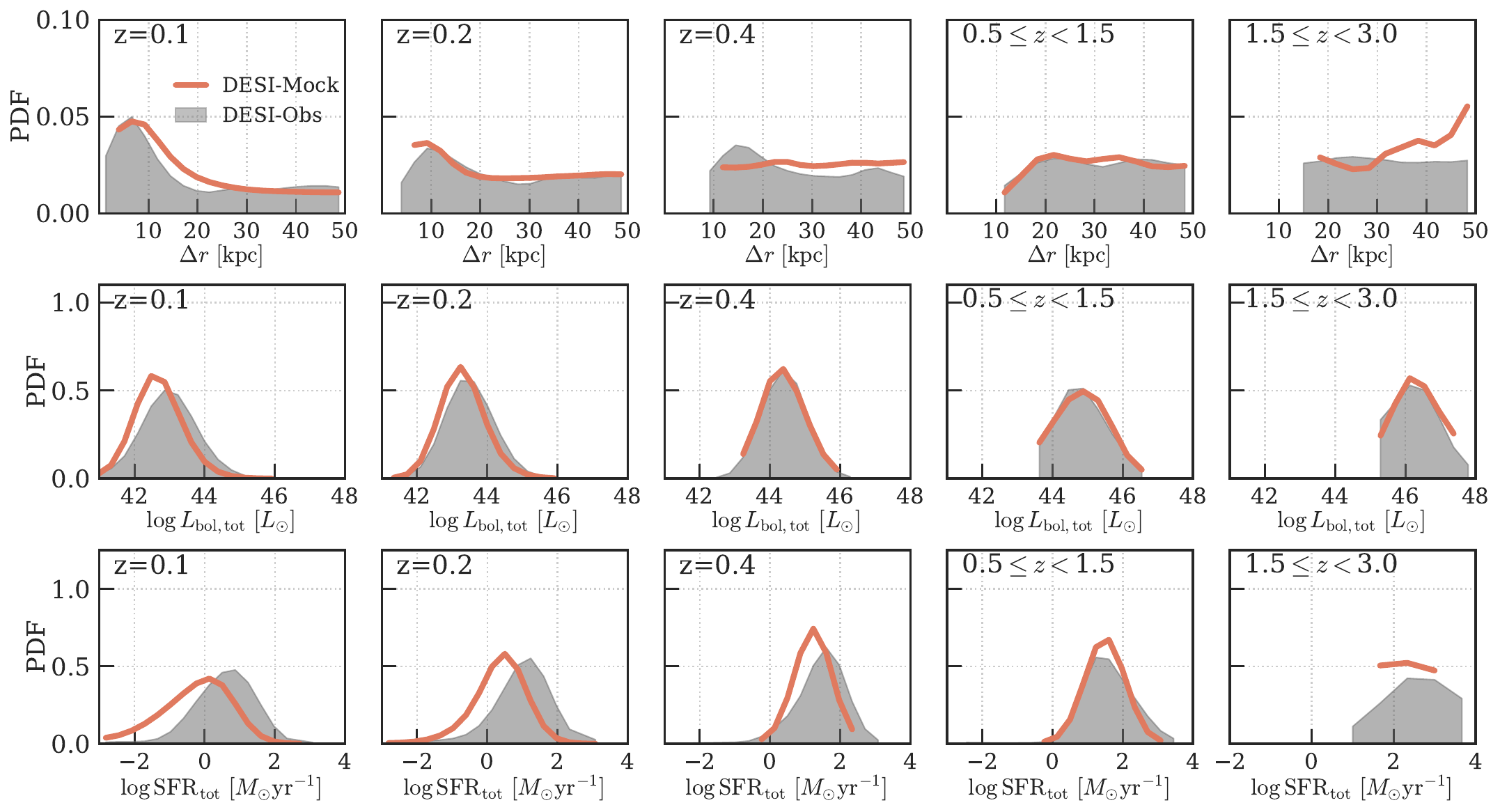}
  \caption{Comparison between the \texttt{DESI-Mock} sample from \texttt{ASTRID} (\textit{red lines}) and the observed DESI dual AGN candidates (\textit{grey shaded}) across five redshift bins. We show comparisons of the projected separation (\textit{top panels}), bolometric luminosities (\textit{middle panels}), and total star-formation rate of the dual AGN host galaxies (\textit{bottom panels}).}
  \label{fig:desi-comparison}
\end{figure*}

\begin{figure*}
\centering
\includegraphics[width=0.95\textwidth]{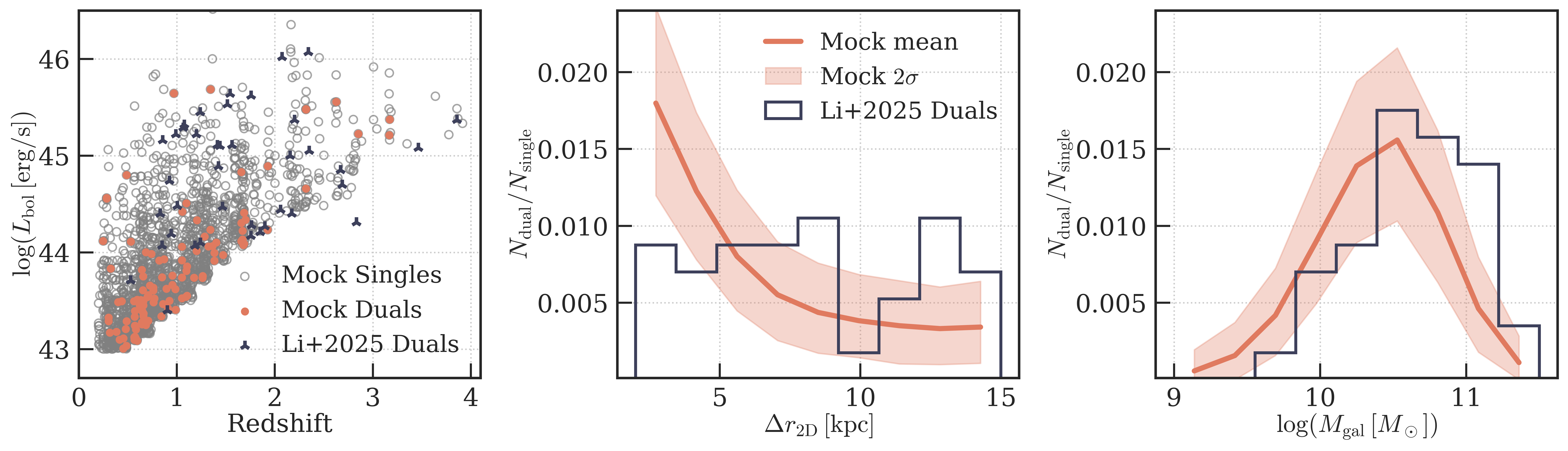}
  \caption{Comparison between realizations of \texttt{COSMOS-Mock} and the observation data from \cite{LiJunyao2025ApJ...986..101L}. \textit{\textbf{Left:}} the redshift and bolometric luminosity in one realization of the \texttt{COSMOS-Mock} sample. We plot the single AGN (\textit{grey circles}) and dual AGN (\textit{red dots}) in this realization. We also show the observed dual candidates from \cite{LiJunyao2025ApJ...986..101L} (\textit{dark blue}). \textit{\textbf{Middle and right:}} projected separation (\textit{middle}) and primary AGN galaxy mass (\textit{right}) distributions of the mock samples, where we  show the mean (\textit{solid red lines}) and 95\% spread (\textit{red shaded}) among 1000 mock light-cone realizations. The distribution of \cite{LiJunyao2025ApJ...986..101L} dual candidates are shown in \textit{dark blue} for comparison. }
  \label{fig:cosmos-comparison}
\end{figure*}

\subsection{Dual AGN and Host Galaxies}
We now turn to a more detailed comparison between our mock dual AGN catalogs and the observation sample, in terms of the separation distribution, galaxy properties, and AGN luminosities.

In Figure \ref{fig:desi-comparison}, we compare the \texttt{DESI-Mock} dual AGN sample from \texttt{ASTRID} with the observed DESI dual AGN candidates from \cite{Dadiani2025} across five redshift bins.
In each bin, we show the distributions of projected separation, bolometric luminosities, and total star-formation rates of the host galaxies, of the simulated and observation samples.
The separation distributions and bolometric luminosities show good agreement between the simulation and observation samples across all redshift bins.
In particular, both the \texttt{DESI-Mock} and the observed samples show an increased dual fraction at close separations of $\Delta r_{\rm 2D} < 10\,{\rm kpc}$, at redshifts where the DESI spatial resolution allows.
The luminosity distribution of both AGN matches almost perfectly between the two samples.

The star-formation rates show good agreement at $z>0.5$ (e.g. for the QSO and ELG pairs). 
Towards lower redshifts, \texttt{DESI-Mock} duals show systematicallly lower SFRs than the observed pairs, especially for the BGS and secondary target samples at $z<0.2$.
A better match to the DESI low-z target (such as applying luminosity and color cuts) may improve the agreement, but the DESI low-z contains significant secondary targets which are not characterized as well as the main survey targets, so we leave a more detailed comparison to future work.

In Figure \ref{fig:cosmos-comparison}, we compare the \texttt{COSMOS-Mock} dual AGN sample with the observed dual candidates from \cite{LiJunyao2025ApJ...986..101L}.
Due to the limited sample size and the uniform selection in separation, we do not bin the data in redshift, but instead compare the overall distributions.
In the left panel, we show one realization of the \texttt{COSMOS-Mock} sample (both the underlying single AGN and the dual AGN) in the redshift-luminosity plane, along with the observed dual candidates.
With the luminosity cut, our mock sample contains a slightly higher fraction of $z<1$ AGN than the observed sample, but overall the redshift and luminosity distributions are similar.

The middle panel of Figure \ref{fig:cosmos-comparison} shows the projected separation distributions of the \texttt{COSMOS-Mock} and the observed dual AGN, including a $95\%$ interval from 1000 light-cone realizations of the mock sample.
There is a slight tension between the simulation and observed separation distributions, where the simulated sample shows a higher dual fraction at $\Delta r_{\rm 2D} < 3\,{\rm kpc}$ than the observed population.
Indeed, \cite{LiJunyao2025ApJ...986..101L} pointed out that there is a lack of dual AGN of close-separation in their sample compared to theoretical expectations and that they found no dual AGN with $\Delta r_{\rm 2D} < 2\,{\rm kpc}$ despite the high spatial resolution of JWST.
Here we quantify the tension at the $> 2\sigma$ level, so the discrepancy may be of physical origin instead of statistical fluctuation.
This comparison indicate that there could still be small-separation duals hidden from current observations, potentially due to the increased obscuration at small-separations \citep[e.g.][]{Chen2023}.
We will discuss in more detail the underlying physical origin of the small-scale peak seen in the simulated sample, and its implications for dual AGN triggering and MBH dynamics in the next section.

In the right panel of Figure \ref{fig:cosmos-comparison}, we compare the stellar mass distribution of the primary AGN of the host galaxy between the mock and observed dual AGN samples.
There is a good agreement between the two samples, both peaks at $M_{\star} \sim 10^{10.5}\,M_\odot$.
There are slightly more high-mass hosts with $M_{\star} > 10^{11}\,M_\odot$ in the observed sample than in the mock sample. 
This is attributed to the higher fraction of low-redshift, low-luminosity AGN in the mock sample, which tend to live in lower-mass hosts.

Overall, we see that after a careful matching of the selection criteria with each observation, the dual AGN population predicted in \texttt{ASTRID} shows reasonable agreement with the key statistical properties of the observed dual AGN samples, including their dual fraction, separation distribution, AGN luminosities, and host galaxy properties.
We will then use this validated mock dual AGN sample to further investigate the physical properties of dual AGN and their implications for MBH binary populations and GW source populations in the next sections.

\subsection{Separation and Dynamical Friction Timescale}
\label{Sec:dr}
\begin{figure}
\centering
\includegraphics[width=0.48\textwidth]{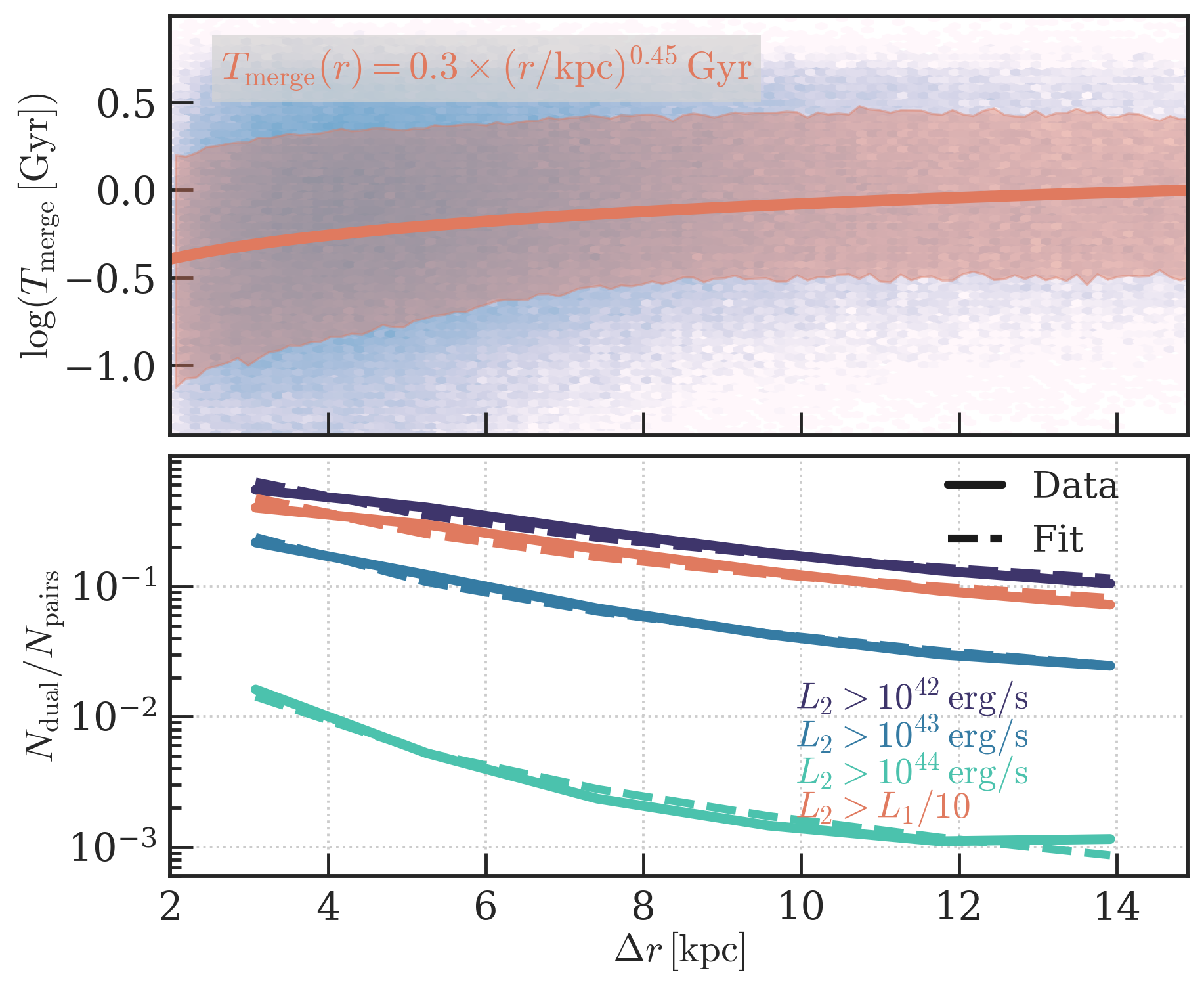}
  \caption{\textit{\textbf{Top:}} Merging timescale as a function of dual AGN separation for all the \texttt{COSMOS-Mock} duals. The background blue histogram shows the distribution of all data. The green shaded region encapsulate the middle $68\%$ of $T_{\rm merge}$ distribution, and the orange curve shows the power-law fit to the data distribution. \textit{\textbf{Bottom:}} companion activation fraction, for companion MBHs ($M_{\rm comp}>10^6\,M_\odot$) to the single AGN in \texttt{COSMOS-Mock} and various luminosity cut of the secondary. We provide power-law fit to the companion activation fraction $f_{\rm dual} = f_0\,(\Delta r/3\,{\rm kpc})^{\alpha}$ in Table \ref{tab:dual-active} and show the fittings in \textit{dashed lines}. }
  \label{fig:dr-component}
\end{figure}

\begin{table}[h!]
\centering
\begin{tabular}{lcc}
\hline
\textbf{$L_{\rm bol, 2}$} & \textbf{$f_0$} & \textbf{$\alpha$} \\
\hline
 $> 10^{42}\,{\rm erg/s}$ & 0.66 & -1.1 \\
$> 10^{43}\,{\rm erg/s}$ & 0.25 & -1.5 \\
$> 10^{44}\,{\rm erg/s}$ & 0.015 & -1.8 \\
$ > L_{\rm bol, 1} / 10$ & 0.49 & -1.2 \\
\hline
\end{tabular}
\caption{Fitting function for the secondary AGN activation timescale show in the bottom panel of Figure \ref{fig:dr-component}. We show the fits for three fixed values of the minimum secondary AGN luminosity, and a case there the secondary luminosity is at least 1/10 of the primary. We take the primary AGN luminosity to be the same as the COSMOS-Mock sample.}
\label{tab:dual-active}
\end{table}

\begin{figure}
\centering
\includegraphics[width=0.48\textwidth]{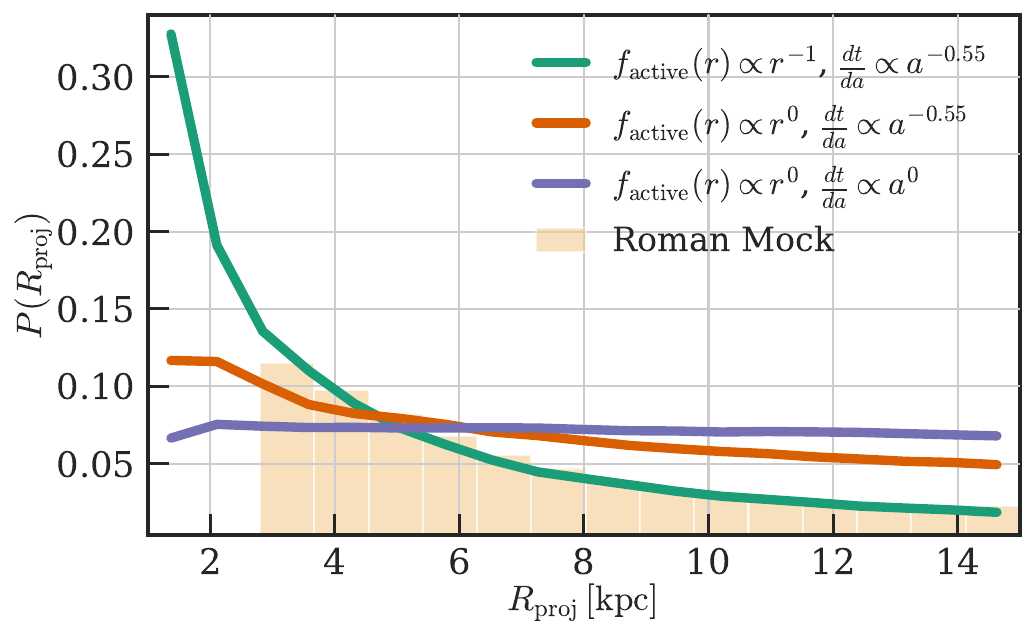}
  \caption{Dual AGN projected separation distribution assuming different secondary activation fraction and sinking timescales following Equation \ref{eq:dr-prob}. We show the model representing the simulation results in Figure \ref{fig:dr-component}: $f_{\rm dual}\propto r^{-1}$ and $\frac{da}{dt}\propto a^{-0.55}$(\textit{green line}), compared with the simulated \texttt{Roman-Mock} sample (\textit{yellow histogram}). The analytical model perfectly reproduces the distribution in the simulation on resolved scales. We also show the distribution assuming separation-independent secondary activation with our fitted sinking timescale (\textit{dark orange}). The \textit{purple line} represent the case when both secondary activation and sinking timescale is independent of pair separation (the latter follows from \cite{Yu2002}). A large sample of dual AGN resolved down to $\Delta r < 2\,{\rm kpc}$ will discriminate between the different physical scenarios.}
  \label{fig:dr-monte-carlo}
\end{figure}

Before diving into the merger and binary populations descended from dual AGN, we spend some time characterizing the spatial distribution of dual AGN and the underlying physical processes shaping it, as this is one of the key differences between the COSMOS-Web observation sample and the simulation predictions.
The spatial distribution of dual AGN contains information about the simultaneous triggering of AGN during galaxy mergers, as well as the dynamical friction timescale of sinking MBH pairs following galaxy mergers.
In observed dual AGN samples, there is often a degeneracy between the effects of dynamical friction and AGN activation.
While in simulations, we can disentangle these two effects and their contributions to the dual AGN separation distribution.

The observed distribution of the projected separation $R$ of all dual AGN can be calculated by integrating the evolution in the semi-major axis $a$ and the true anomaly $\nu$, taking into account the orbital eccentricity $e$ and a projection from 3D separation $r$:
\begin{equation}
\label{eq:dr-prob}
    P(R) \propto \int_{a_0}^{a_1} da \frac{dt}{da}(a) \frac{1}{T(a)}\int_0^{2\pi} d\nu \frac{dt}{d\nu}f(r)\, P(R|r),
\end{equation}
where $\frac{dt}{da}$ is the orbital evolution timescale, $T(a)\propto a^{3/2}$ is the orbital period, $\frac{dt}{d\nu} \propto \frac{r^2}{\sqrt{a(1-e^2)}}$ is the time spent at true anomaly $\nu$, $f(r)$ is the dual activation fraction as a function of separation, and $P(R|r) = \frac{R}{r^2\sqrt{1-(R/r)^2}}$ is the projection kernel.
To minimize the scaling relations needed, we assume that the MBHs follow Keplerian orbits with non-evolving eccentricity, but eccentricity evolution can be easily added to the above expression.
We see from the above expression that building a model for the projected separation distribution requires modeling of the sinking timescale $\frac{dt}{da}$ and the dual activation fraction $f(r)$.

In the top panel of Figure \ref{fig:dr-component}, we show the merging timescale $T_{\rm merge}$ as a function of dual AGN separation for all the \texttt{COSMOS-Mock} duals.
Albeit the large scatter, we can fit a power-law relation to the data, with $T_{\rm merge} = 0.3\,{\rm Gyr}\,(\Delta r / 1\,{\rm kpc})^{0.45}$.
This scaling is shallower than the classic Chandrasekhar dynamical friction timescale scaling of $T_{\rm DF} \propto r^2$, which assumes an isothermal sphere background and a constant Coulomb logarithm \citep{Chandrasekhar1943, Binney2008}.
On the other hand, it is consistent with more recent numerical studies of dynamical friction in galaxy mergers, which find a shallower scaling of $T_{\rm DF} \propto r^{0.5-1}$ due to the complex background potential and time-evolving Coulomb logarithm \citep[e.g.][]{Yu2002,Boylan-Kolchin2008MNRAS.383...93B, Dosopoulou2017}.

The bottom panel of Figure \ref{fig:dr-component} shows the companion activation fraction as a function of dual AGN separation, for various luminosity thresholds of the secondary MBH.
We see a clear trend of higher activation fraction at smaller separations, which can be well described by a power-law relation $f_{\rm dual} = f_0\,(\Delta r/3\,{\rm kpc})^{\alpha}$, with the fitting parameters summarized in Table \ref{tab:dual-active}.
The increasing activation fraction towards small separations arises because in \texttt{ASTRID}, galaxy mergers drive gas inflows towards the nuclear regions and activate both MBHs simultaneously.
The secondary activation mechanism may differ if the dual AGN are triggered by secular processes instead of galaxy mergers, or if the gas distribution in merging galaxies are more clumpy and turbulent than in the simulation.

With the above fittings for the sinking timescale and companion activation fraction, we can now build a simple analytical model for the dual AGN separation distribution using Equation \ref{eq:dr-prob}.
In Figure \ref{fig:dr-monte-carlo}, we show the dual AGN projected separation distribution assuming different secondary activation fraction and sinking timescales.
The model representing the simulation results is shown in green, which assumes $f_{\rm dual}\propto r^{-1}$ and $\frac{dt}{da}\propto a^{-0.55}$.
We see that the fitted model results in a peaked separation distribution at small scales.
We show the separation distribution from the \texttt{Roman-Mock} sample in yellow for comparison, and find that despite the scatter in the sinking timescale and secondary activation fraction, the analytical model perfectly reproduces the distribution in the simulation on resolved scales.

Together with the simulation-fitted model, we have two alternatives: a model with separation-independent secondary activation but the same sinking timescale, and a model with both secondary activation and sinking timescale independent of pair separation. 
Both scenarios result in much flatter separation distributions, with no peak at small separations.
We see that with the dual sample from the upcoming Roman survey, which is expected to resolve dual AGN down to $\Delta r < 1\,{\rm kpc}$ and has a large survey area, we will be able to discriminate between these different physical scenarios and constrain the underlying MBH dynamics and dual AGN triggering mechanisms.

\section{From Dual AGN to MBH Mergers}
\label{sec:merger}

\begin{figure*}
\centering
\includegraphics[width=0.49\textwidth]{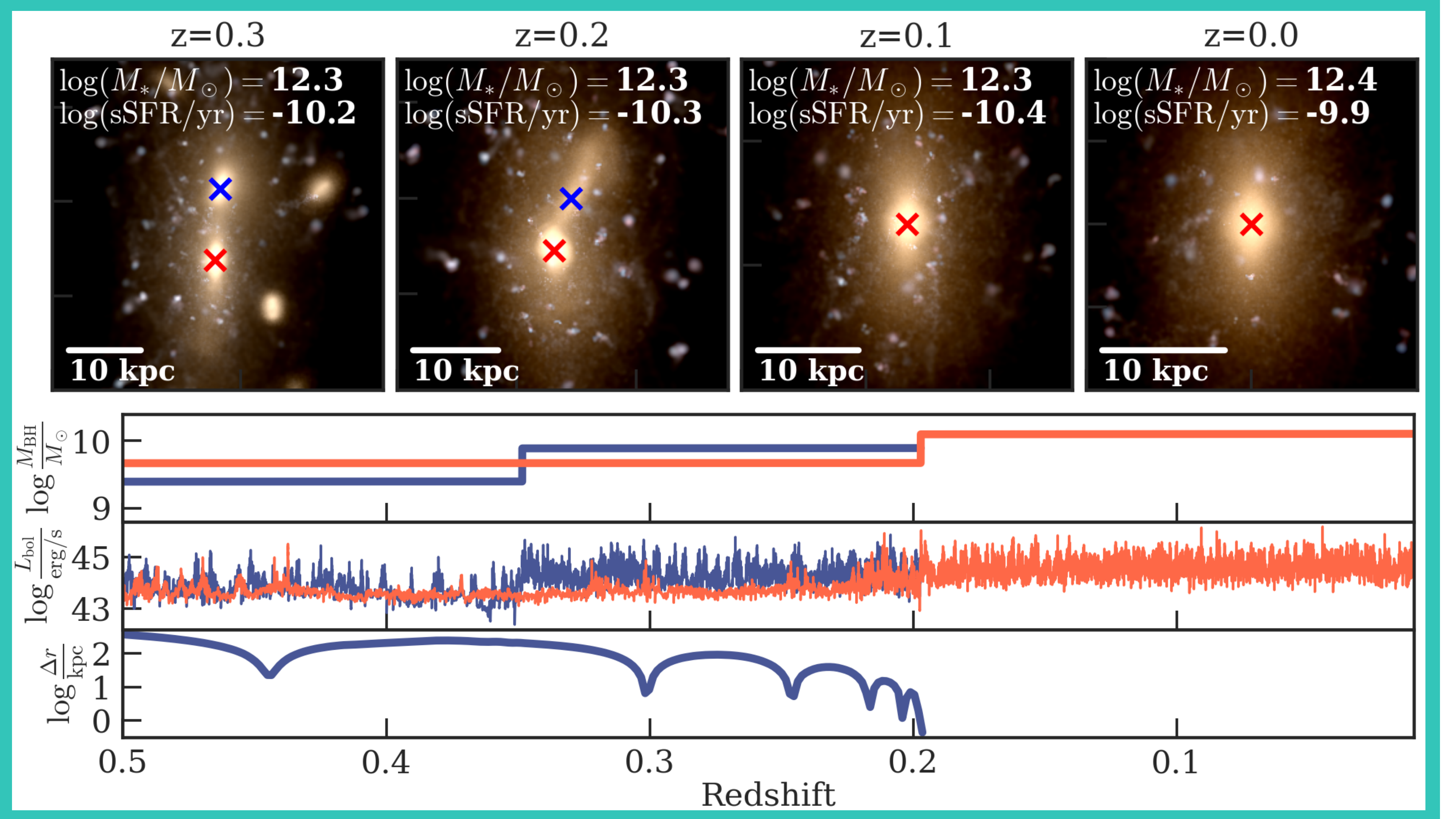}
\includegraphics[width=0.49\textwidth]{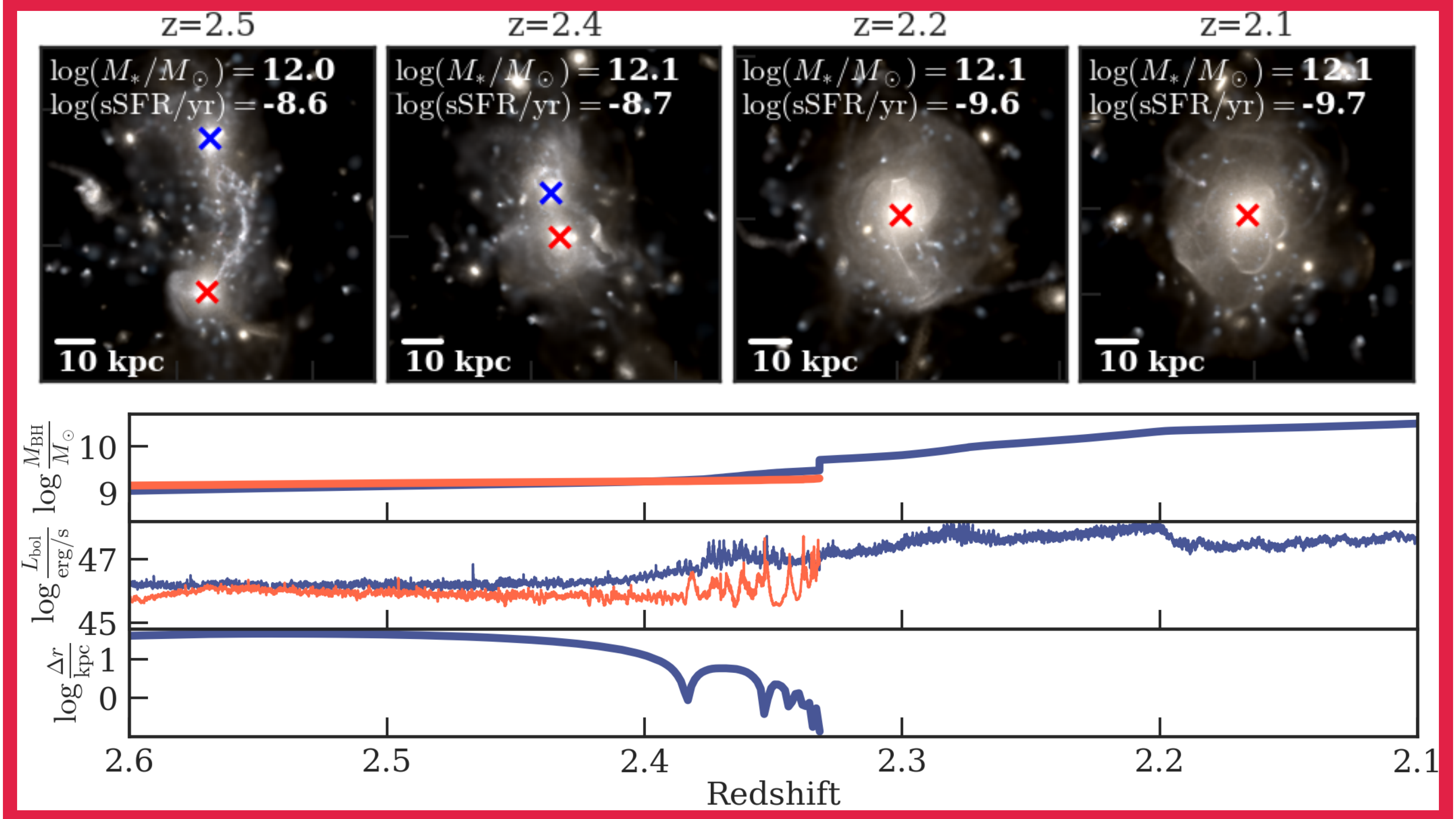}
\includegraphics[width=0.49\textwidth]{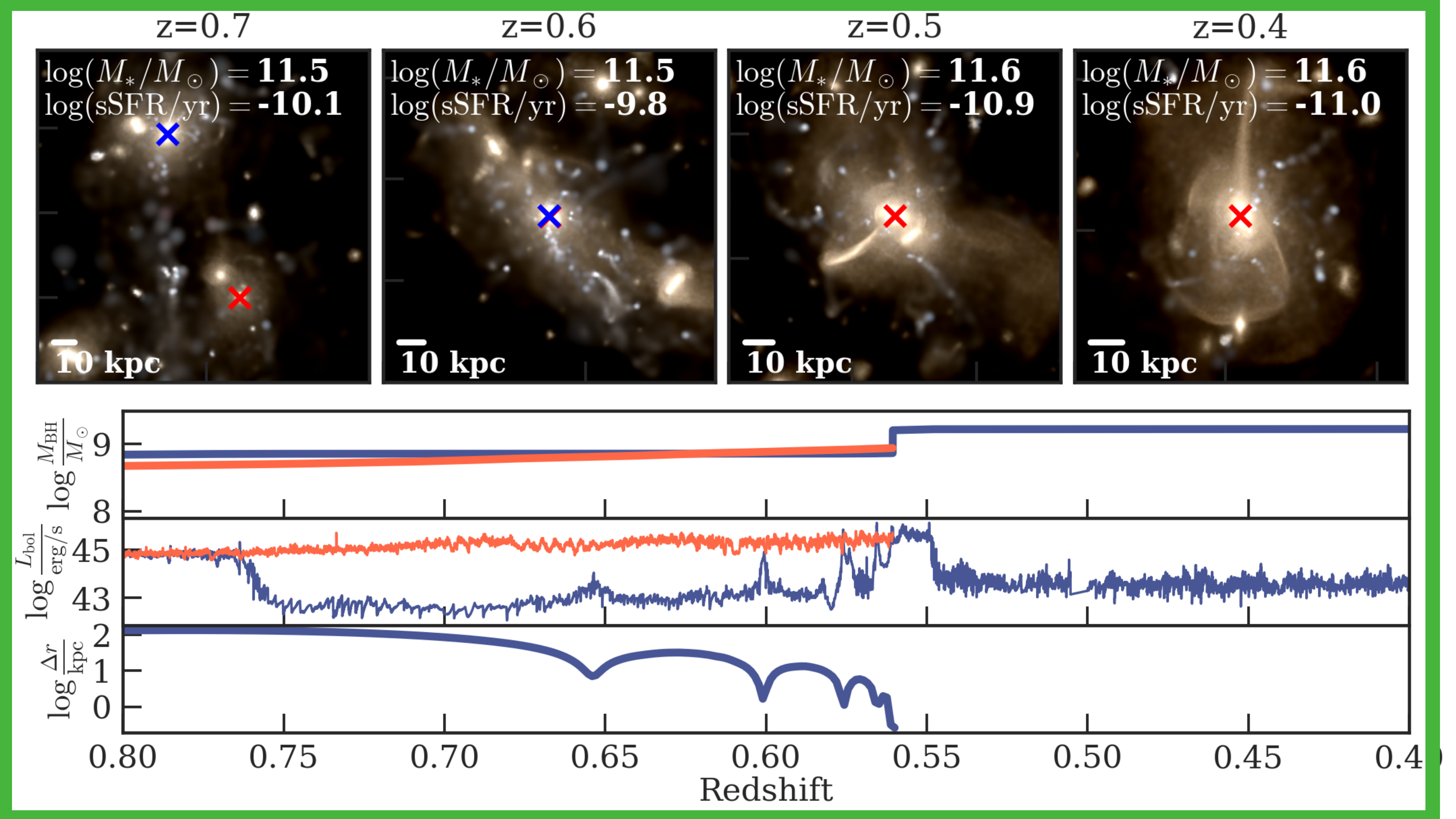}
\includegraphics[width=0.49\textwidth]{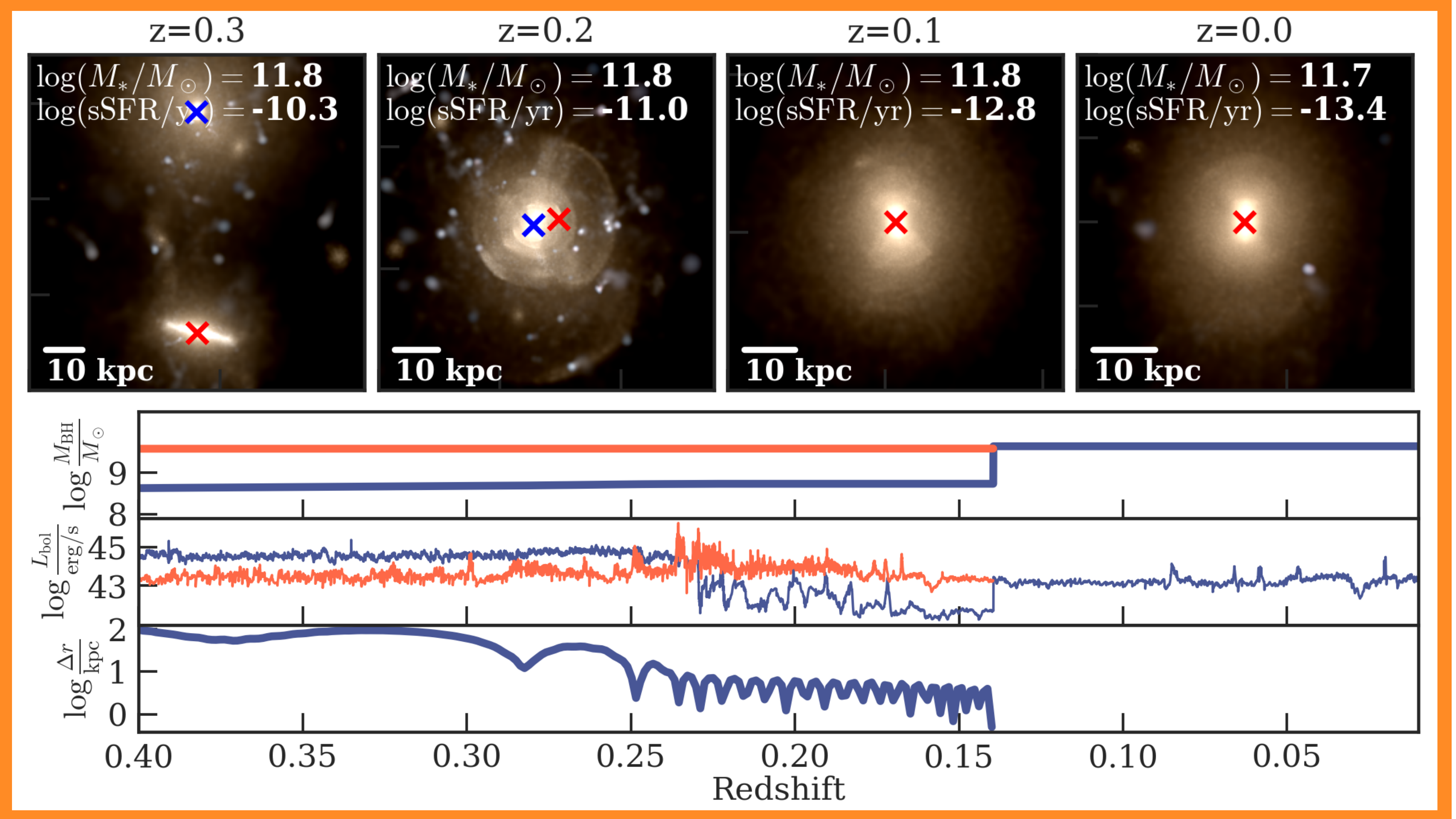}
\caption{Four example dual AGN systems and their evolution to MBH mergers. \textit{Upper left:} a low-redshift dual AGN that will evolve into a resolvable PTA single source (see Figure \ref{fig:merger-mz}). \textit{Upper right:} a high-redshift dual quasar system that evolves into one of the brightest quasars in the simulation ($L_{\rm bol} > 3\times 10^{47}\,{\rm erg/s}$). The tidal feature in the host galaxy is still prominent a few hundred Myrs after the simulation merger. \textit{Lower left:} a dual AGN system moving into the ''green-valley" region after the merger of MBHs. This system is in a very complex environment with many satellites, and therefore, the AGN becomes inactive while the galaxy remains star-forming. \textit{Lower right:} a dual AGN evolving into a massive, red elliptical at $z\sim 0$.}
  \label{fig:evol-example}
\end{figure*}

\begin{figure*}
\centering
\includegraphics[width=0.98\textwidth]{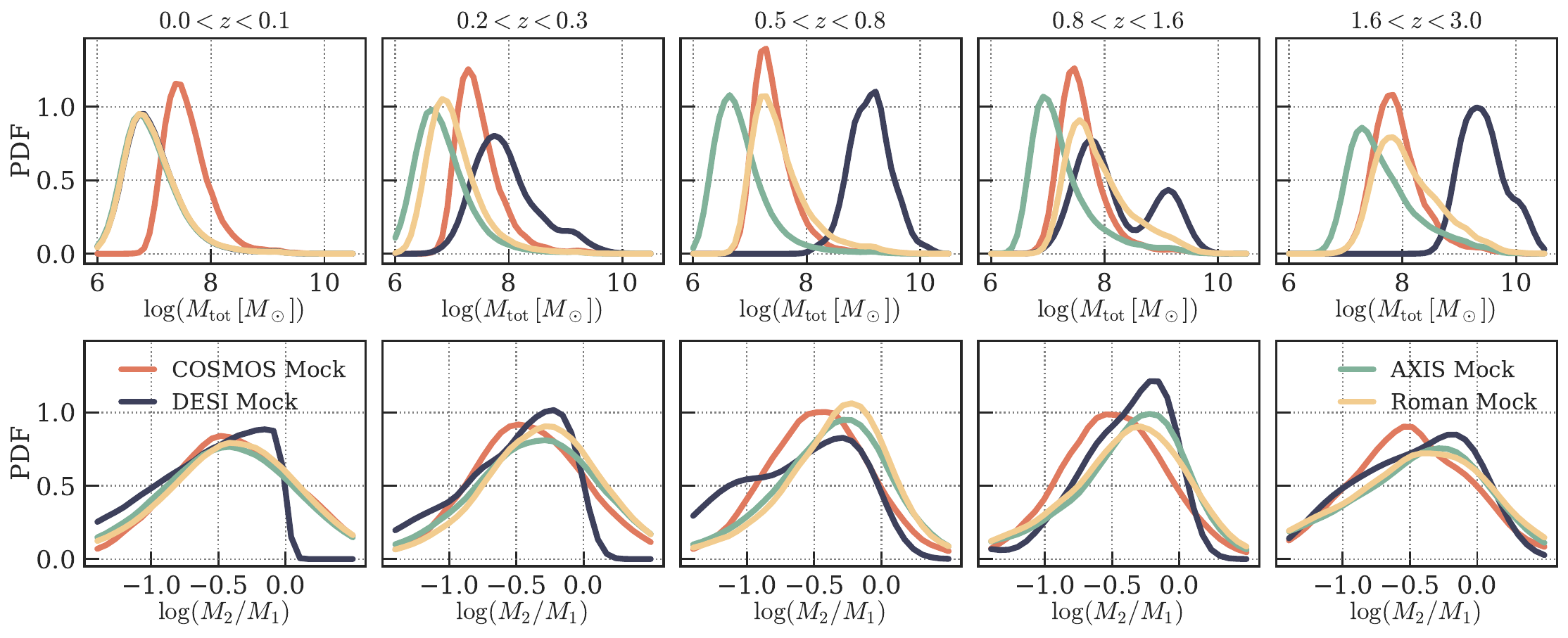}
  \caption{\textit{\textbf{Top:} } Total BH mass distribution of dual AGN in the four mock samples: \texttt{Cosmos-Mock} (\textit{red}), \texttt{DESI-Mock} (\textit{dark blue}), \texttt{AXIS-Mock} (\textit{green}) and \texttt{Roman-Mock} (\textit{yellow}). We show the distribution in five redshift bins (from left to right): $0<z<0.1$, $0.2<z<0.3$, $0.5<z<0.8$, $0.8<z<1.6$, and $1.6<z<3.0$.
  \textit{\textbf{Bottom:}} mass ratio distribution of dual AGN in the mock samples. The mass ratio here is the ratio between the mass of the fainter AGN and that of the brighter AGN, so the value can exceed one.}
  \label{fig:mass_distribution}
\end{figure*}

Dual AGN are direct progenitors of gravitational wave events, and are thus of great interest for multi-messenger studies of the MBH population across cosmic time. 
By studying the host galaxy properties of the kpc-scale dual AGN, we gain better insight into the EM signature of GW events. 
In this section, we use simulation to link dual AGN with MBH binaries/merger events and evaluate the contribution of dual AGN remnants to the GW sources in the LISA band as well as the gravitational-wave background in the PTA band.

We begin by illustrating the evolution from dual AGN to SMBH mergers in Figure \ref{fig:evol-example}, where we show four simulated dual AGN along with the host galaxies.
The first column of the top panels in each system shows the MBH and host galaxies when the system is captured as a dual AGN in one of the four mock catalogs.
We then forward model them in the simulation until $\sim 0.5-1\,{\rm Gyr}$ after the numerical merger.
We see the diverse evolution paths among duals: some remain star-forming due to abundant gas or infalling satellites, while others go through rapid quenching immediately after the galaxy mergers.
We also note that some systems, such as the green panel, still show strong tidal features $1\,{\rm Gyr}$ after the simulation merger.
Future work can explore in more detail the identification of binary hosts from morphological features.

\subsection{MBH masses probed by dual AGN}

LISA and PTA are sensitive to different mass and redshift ranges of MBH binaries.
LISA is most sensitive to binaries with total masses between $10^4-10^7\,M_\odot$ at redshifts $z\sim 1-5$ \citep[e.g.][]{Amaro-Seoane2017arXiv170200786A}, while PTA experiments probe the stochastic gravitational-wave background produced by the in-spiral of more massive binaries ($10^7-10^{10}\,M_\odot$) at lower redshifts ($z<2$) \citep[e.g.][]{NG15_binary, EPTA2023A&A...678A..50E}.
Here, we examine the evolution in the mass distribution probed by each dual AGN search to help understand the overlap with GW searches and the dual AGN merging timescales.

In the top panel of Figure \ref{fig:mass_distribution}, we show the mass distributions of the dual AGN in the four mock samples in five redshift bins.
\texttt{COSMOS-Mock} has the weakest redshift dependence and almost always probes dual AGN with masses between $10^7-10^8\,M_\odot$ across all redshifts.
\texttt{DESI-Mock} dual AGN show a strong redshift evolution in mass distribution, due to the sample selection targeting various DESI tracer populations at different redshifts.
At $z>1.6$, DESI duals mainly consist of very massive MBHs ($>10^9\,M_\odot$) in luminous quasars.
The bi-modal mass distribution at $0.8<z<1.6$ reflects the two DESI target populations: luminous quasars with $M_{\rm BH}\sim 10^9\,M_\odot$ and less massive AGN in ELGs with $M_{\rm BH}\sim 10^8\,M_\odot$.
AXIS duals mainly probe lower-mass MBHs ($10^6-10^7\,M_\odot$) at all redshifts due to their high sensitivity in the X-ray band.
At low redshifts ($z<0.3$), DESI, AXIS and Roman probe similar mass ranges between $10^6-10^8\,M_\odot$, which are the prime LISA source progenitors.

The bottom panel of Figure \ref{fig:mass_distribution} shows the distribution of the dual AGN mass ratio in the four mock samples.
We see that across all redshifts, the dual AGN by these observational studies probes major mergers with mass ratios between $1$ and $1:3$.
This would allow for a relatively efficient sinking of the secondary MBH after galaxy mergers, and thus a high merger fraction of dual AGN.
The \texttt{COSMOS-Mock} sample has a slightly lower mass ratio peaking at $1:3$. 
This is because this sample is dominated by small-separation duals, where the secondary has a higher Eddington ratio as it approaches the primary host galaxy.
We also note that across all redshifts, there is a relatively small fraction ($\sim 20\%$) of dual AGN in which the brighter AGN is the less massive MBH.

\subsection{Dual AGN Merging timescale}
\begin{figure}
\centering
\includegraphics[width=0.46\textwidth]{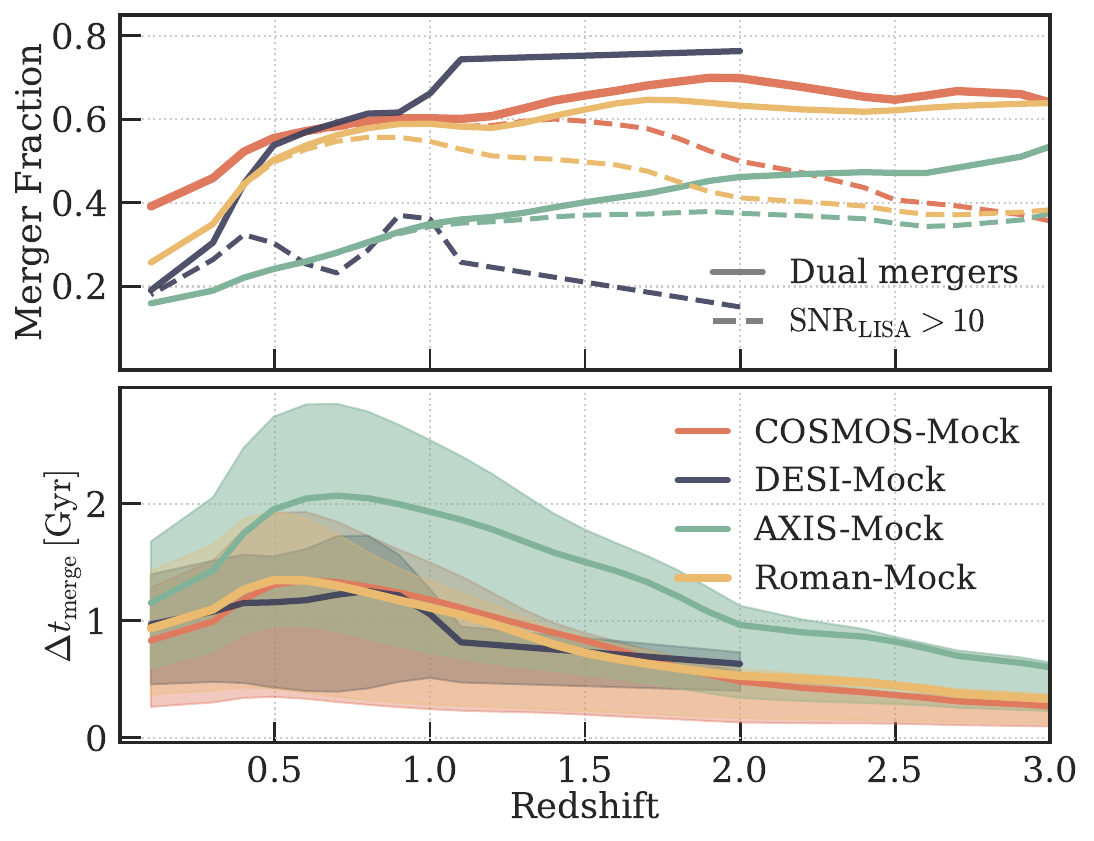}
  \caption{\textit{\textbf{Top:}} The fraction of dual AGN that merged by $z=0$ in each mock samples (\textit{solid lines}). The fraction of LISA-detectable sources (${\rm SNR}>10$) is shown in \textit{dashed lines}. \textit{\textbf{Bottom}}: Dual AGN merging timescale for each mock dual AGN sample at different redshifts. The \textit{solid lines} show the mean merging time and the \textit{shaded regions} represent the central 90\% of the distribution. }
  \label{fig:merger_frac}
\end{figure}

We now investigate the relation between dual AGN and MBH mergers by tracing whether and when they will evolve into MBH mergers in the simulation.
We count both direct mergers between the dual AGN and mergers between their descendants if they merge with other BHs before each other.
We note that the ``merger" refers to simulation mergers, which happen at the late stage of the dynamical friction-dominated orbital decay, when the MBH pairs are separated by $\sim 1\,{\rm kpc}$.
For this work, we do not fully model the binary evolution timescale as in, e.g. \citet{Chen2022}.
We assume a binary formation timescale of $0-500\,{\rm Myr}$ after the simulation merger, motivated by the short merging timescale in massive star-forming galaxies found in \cite{Liao2024MNRAS.528.5080L}. 
The galaxy and AGN properties of the binary/merger events are then extracted in the closest snapshot within $500\,{\rm Myr}$ of the simulation merger.

In the top panel of Figure \ref{fig:merger_frac}, we plot the fraction of dual AGN that would evolve into an MBH merger as a function of redshift.
At high redshifts ($z>1$), duals in COSMOS, DESI, and Roman have a high merger fraction of $\sim 60-80\%$ (for DESI we stacked the samples at $z>1.6$ to get the fraction around $z=2$ due to the low number density).
For DESI, this is mainly because the duals at this redshift range are the most massive quasars with a high density of star-forming gas in the host galaxy, which enables efficient dynamical friction and rapid MBH sinking.
The high merger fraction for COSMOS and Roman is a consequence of the small separations probed by these two samples.
AXIS duals have the lowest BH masses and relatively large separation, and thus the lowest merger fraction of $\sim 30-40\%$ across all redshifts.
Our result is in line with the $30-70\%$ merger fraction found in \cite{Volonteri2022}, using the Horizon-AGN simulation also with subgrid dynamical friction.

We then calculate the fraction of dual AGN that would evolve into LISA-detectable mergers (with ${\rm SNR}>10$ following the method in \citealt{Chen2022}), using the \texttt{BOWIE} package \citep[][]{Katz2020, Katz2023ascl.soft07015K}.
We see that among all dual AGN samples, COSMOS duals have the highest LISA merger fraction of $\sim 50-70\%$ at $z\sim 0.5$, due to the small separations of the dual AGN.
Despite the high merger fraction of the DESI duals at $z>1$, only a small fraction ($20\%$) of them would be LISA-detectable due to their high masses ($>10^8\,M_\odot$).

In the bottom panel of Figure \ref{fig:merger_frac}, we show the merging timescale distribution of dual AGN in the four mock samples at different redshifts.
COSMOS, DESI and Roman duals have very similar merger timescale distributions and evolution, with a median merging timescale of $\sim 0.3-0.5\,{\rm Gyr}$ at $z>1$ and $\sim 1\,{\rm Gyr}$ at $z<0.5$.
AXIS duals have a longer merging time scale of $\sim 1-3\,{\rm Gyr}$, due to their lower masses and larger separations.
The decline in the median merger timescale after $z=0.5$ arises due to the limited time until $z=0$ for the merger to occur.
The mergers after $z=0.5$ select only the duals that merge quickly enough.

Overall, we see that dual AGN are robust tracers  of mergers, and thus by studying the host galaxies of dual AGN, we can narrow down the galaxies that potentially host MBH binaries.
Furthermore, dual AGN evolve relatively quickly to mergers due to the abundance of gas and star-formation in dual hosts.
It is thus also possible that the host galaxy still retains merger signatures at the time of MBH binary formation.

\subsection{Dual AGN as GW Sources in LISA and PTA}
\begin{figure}
\centering
\includegraphics[width=0.44\textwidth]{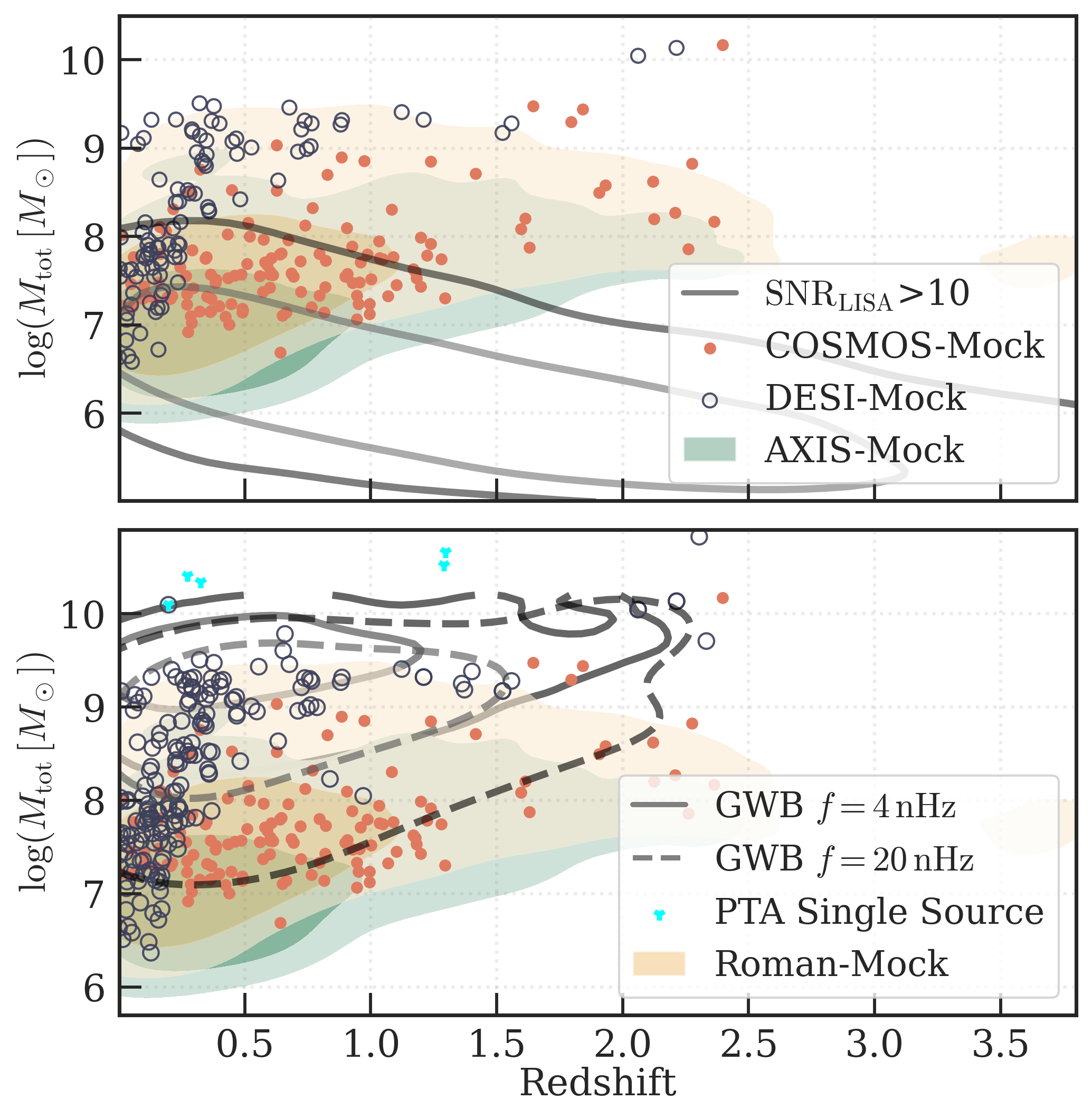}
  \caption{Redshift and mass distribution of dual AGN in four mock samples. We show mergers from one realization of the \texttt{COSMOS-Mock} and \texttt{DESI-Mock} ($8\,{\rm deg}^2$) sample in \textit{red} and \textit{dark blue} dots, respectively. Mergers from AXIS (\textit{green}) and Roman (\textit{yellow}) duals are shown in shaded contours. In the \textit{\textbf{top panel}}, we overlay the dual merger distributions with the LISA sources in \texttt{ASTRID} (\textit{grey solid contours}). In the \textit{\textbf{bottom panel}}, we show the same merger sample (but a different realization for DESI) overlaid with the GWB source distribution.}
  \label{fig:merger-mz}
\end{figure}

We have seen in the previous section that a significant fraction of dual AGN would evolve into MBH mergers by $z=0$, on a timescale of $\sim 0.5-1\,{\rm Gyr}$.
We now investigate the contribution of dual AGN remnants to the MBH mergers detectable by LISA and the gravitational-wave background in the PTA band.

\subsubsection{LISA Sources}
In the top panel of Figure \ref{fig:merger-mz}, we show the mass and redshift distribution of dual AGN mergers from each mock catalog, at the time of the simulation merger, overlaid with the LISA sources in \texttt{ASTRID} \citep[][]{Wang2025arXiv250324304W}.
For the \texttt{COSMOS-Mock} sample, we show one example light-cone realization covering $0.2\,{\rm deg}^2$, and for the \texttt{DESI-Mock} sample we show one realization covering $8\,{\rm deg}^2$ (this is a small fraction of the DESI DR1 footprint).
The AXIS and Roman mergers are shown in shaded contours, with the dark and light regions enclosing the central 68\% and 95\% of the distribution, respectively.

At high redshifts, the dual AGN samples probe mainly the luminous duals due to the flux limits of the surveys, whereas LISA is most sensitive to lower-mass MBH mergers (due to the V-shaped sensitivity curve of LISA, massive mergers are below the left side of the sensitivity curve at high redshifts).
From \cite{Chen2022} and \cite{Wang2025arXiv250324304W}, we have shown that despite the longer merging timescale of MBH seeds, seed-mass mergers ($M_{\rm tot}<10^6\,M_\odot$) still dominate LISA sources in all redshifts. 
Thus, there is little or no overlap between dual AGN mergers and LISA sources at $z>1.5$.
At low redshifts, Roman and AXIS dual mergers start entering the LISA band because of their sensitivity to fainter and low-mass MBHs.
Between $0<z<0.5$, there is a significant overlap between LISA sources and dual AGN mergers from AXIS and Roman, with a total black hole mass between $10^6-10^7\,M_\odot$.
DESI dual mergers are also promising LISA progenitors at $z<0.2$, while COSMOS dual mergers only probe the high-mass end of LISA sources due to the minimum bolometric luminosity cut at $10^{43}\,{\rm erg/s}$.

\begin{figure}
\centering
\includegraphics[width=0.46\textwidth]{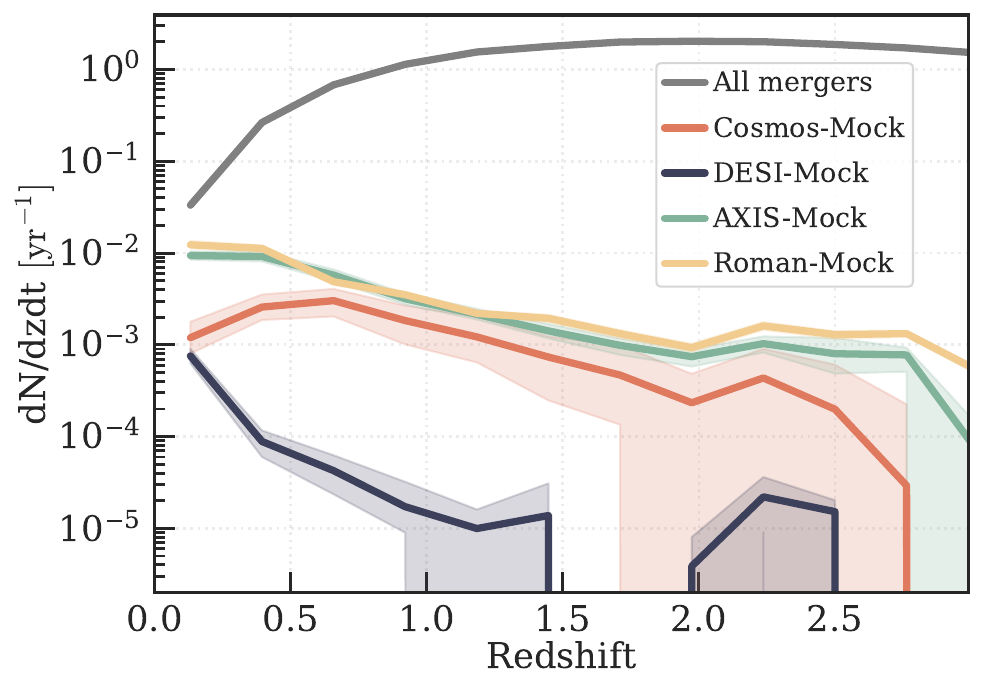}
  \caption{Expected full-sky merger rates from COSMOS (\textit{red}), DESI (\textit{dark blue}), AXIS (\textit{green}) and Roman (\textit{yellow}) dual AGN mock samples, compared with the total MBH merger rate predicted by \texttt{ASTRID} (\textit{grey}). We expect $\sim  10^{-4}\,{\rm yr}^{-1}$ \texttt{DESI-Mock} induced mergers from all directions, out of which only $2\times \sim 10^{-5}\,{\rm yr}^{-1}$ would fall within the DESI DR1 footprint. Mergers from \texttt{Cosmos-Mock} duals are expected at a rate of $0.01\,{\rm yr}^{-1}$ (full-sky), but only $6\times 10^{-8}\,{\rm yr}^{-1}$ of these would be detectable within the COSMOS-Web field.}
  \label{fig:mrate}
\end{figure}

In Figure \ref{fig:mrate}, we calculate the expected full-sky merger rate from dual AGN in the four mock samples, and compare them with the total LISA MBH merger rate predicted by \texttt{ASTRID}.
We note that this does not take into account the survey area of each observation but represents the expected mergers from dual AGN with similar selection criteria across the full sky.
The total LISA merger rate is expected to peak around $z=2$ at a rate of $\sim 3\,{\rm yr}^{-1}$.
The dual AGN mergers peak around a much later epoch around $z=0.3-0.5$, with a total rate ranging from $0.02\,{\rm yr}^{-1}$ for \texttt{AXIS-Mock} duals to $\sim  10^{-4}\,{\rm yr}^{-1}$ for \texttt{DESI-Mock} duals.
This means that the chances of probing a LISA progenitor from current dual AGN surveys are low.
However, AXIS and Roman will provide significant samples of LISA progenitors at $z<0.3$, where $>10\%$ of the LISA sources have a dual AGN progenitor in these two surveys.
If any LISA sources are detected in the local volume, there is a relatively high chance that their dual AGN progenitors could be found in AXIS or Roman surveys.

\begin{figure}
\centering
\includegraphics[width=0.48\textwidth]{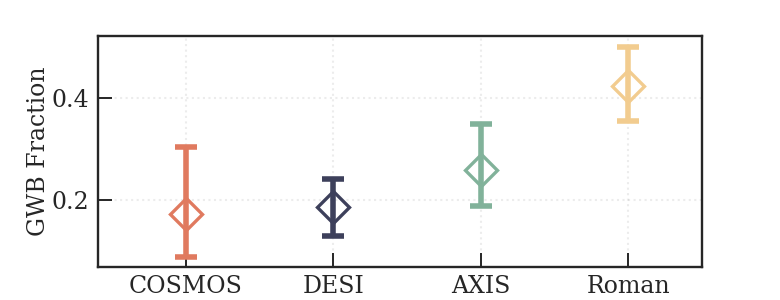}
  \caption{Fraction of the gravitational wave background produced by MBH binaries descended from dual AGN in the \texttt{COSMOS-Mock} (\textit{red}) and \texttt{DESI-Mock} (\textit{dark blue}), AXIS (\textit{green}), and Roman (\textit{yellow}) samples, in the low-frequency bin of the PTA band ($f=0.1\,{\rm yr}^{-1}$).}
  \label{fig:gwb_frac}
\end{figure}

\subsubsection{PTA Sources}
Although current dual AGN observation samples do not constitute many of the LISA-detectable events because they probe the brightest and most massive population among MBH pairs and mergers, they could provide good progenitor samples for lower-frequency GWs from SMBH binaries.
In previous works such as \cite{Padmanabhan2024A&A...684L..15P}, dual AGN observation was used along with the observed GWB to constrain the merging timescale and the simultaneous triggering of AGN in galaxy mergers.
It has also been shown in \cite{Zhou2025ApJ...988L..74Z} that dual AGN remnants could constitute more than $70\%$ of the PTA single source population.
Understanding the connection between dual AGN and the binary population producing the observed GWB in the PTA band enables us to better interpret the observed signal and constrain the MBH population and their dynamical evolution.

In the bottom panel of Figure \ref{fig:merger-mz}, we show the mass and redshift distribution of dual AGN mergers from each mock catalog, overlaid with the MBH binaries producing the GWB in \texttt{ASTRID}.
We generate GWB from the simulated MBH mergers following the method in \cite{Chen2025ApJ...991L..19C} \citep[also see][]{Kelley2017b}, and use the same analytical model for sub-resolution orbital decay timescales (which follows from the constraints in \cite{NG15_binary}). 
The binary distribution is weighted by their $h_s^2$ contribution to the total stochastic GW background, so the distribution peaks around massive binaries despite their low number density \citep[see e.g.][for details on the GWB distribution from the \astrid simulation]{Chen2025ApJ...991L..19C}.

We see that the low-redshift and massive end of the dual AGN mergers overlap significantly with the GWB-producing binary population.
In fact, the GWB from the $z<0.2$ binaries almost all comes from dual AGN descendants.
We also show five PTA foreground sources with high detection probability in cyan stars generated following \citep{Zhou2025ApJ...988L..74Z}.
We see that one of the foreground sources is a direct descendant of a dual AGN in the \texttt{DESI-Mock} sample.

\begin{figure*}
\centering
\includegraphics[width=0.95\textwidth]{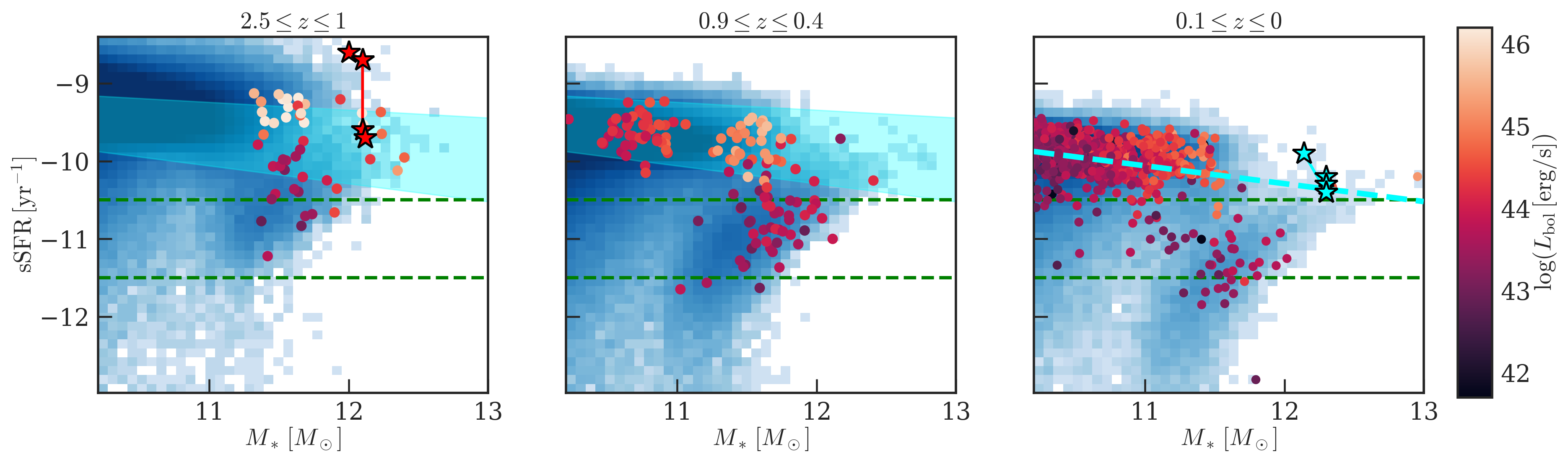}
  \caption{DESI Mock dual AGN merger remnant (red and purple dots) plotted on top of the single AGN population (blue background) in three redshift bins. The \textit{cyan shaded} region in the left and middle panel shows the star-forming main sequence fitted in \citet{Elbaz2007A&A...468...33E} at $z=0$ and $z=1$, and the cyan dashed line in the right panel is the same fit at $z=0$. The region between the two solid dashed encloses the green-valley galaxies \citep[e.g.][]{Salim2014SerAJ.189....1S}. The red and cyan stars show the evolution tracks of two systems in the upper panel of Figure \ref{fig:evol-example} from dual AGN to binary/mergers.}
  \label{fig:binary-desi}
\end{figure*}

\begin{figure*}
\centering
\includegraphics[width=0.95\textwidth]{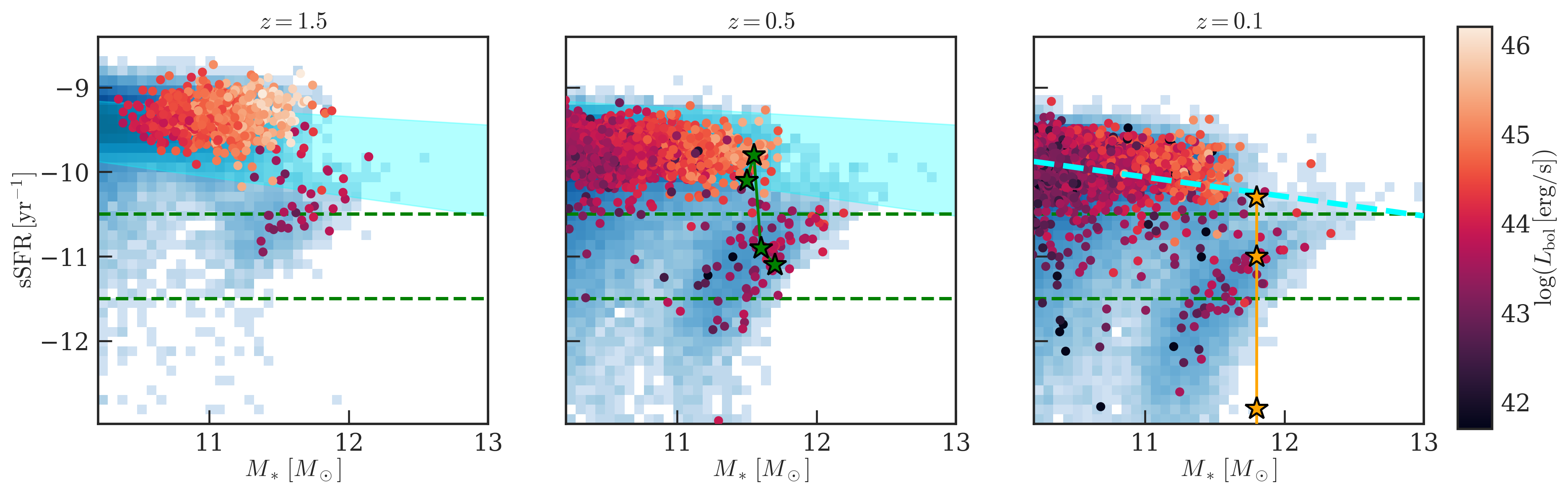}
  \caption{
  Same as Figure \ref{fig:binary-desi} but for merger remnant of the Roman mock dual AGN sample, at three redshifts. The green and orange stars show the evolution tracks of two systems in the lower panel of Figure \ref{fig:evol-example} from dual AGN to binary/mergers.
  }
  \label{fig:binary-roman}
\end{figure*}

Finally, in Figure \ref{fig:gwb_frac}, we show the fraction of GWB produced by dual AGN descendants that have similar properties as targeted by each survey, at a frequency of $f=0.1\,{\rm yr}^{-1}$ in the PTA band.
We find that dual AGN remnants constitute a significant fraction of the total GWB: $\sim 40\%$ for AXIS and Roman duals and $\sim 20\%$ for COSMOS and DESI.
The background missed from the dual AGN population comes from the most massive binaries with $M_{\rm total}>10^9\,M_\odot$ at $0.3<z<1.3$, which have merging timescales too short to be captured by dual AGN searches.
These massive MBH pairs are also already in the radio mode before pairing and are thus elusive to the dual AGN searches.
We thus conclude that dual AGN are reliable tracers of the MBH binary population producing the GWB in the PTA band, and future dual AGN surveys will provide important constraints on the MBH binary population.

\subsection{Where to Find Dual AGN Descendants?}

In the previous section, we have shown that most dual AGN eventually evolve into SMBH binaries that dominate the stochastic gravitational-wave background in the PTA band.
With evidence for this background now established, PTAs are expected to detect individual SMBHBs whose signals rise above it  \citep[e.g.][]{Rosado2015MNRAS.451.2417R, Mingarelli2017NatAs...1..886M, Kelley2019BAAS...51c.490K}.
Because SMBHBs can emit both strong gravitational and bright electromagnetic signals, they are prime multi-messenger targets \citep[e.g.][]{Sesana2012MNRAS.420..860S, Charisi2022MNRAS.510.5929C}, enabling constraints on binary–environment interactions, accretion physics, and SMBH–galaxy co-evolution.
In this section, we explore the galaxy-AGN parameter space and identify regions where SMBH binaries evolved from dual AGN are most likely to be found.

Figure \ref{fig:binary-desi} and \ref{fig:binary-roman} show the host galaxy properties and AGN luminosities of the merger remnants of dual AGN in the \texttt{DESI-Mock} and \texttt{Roman-Mock} samples, respectively.
We only show these two samples because of their large survey area and thus a higher chance of finding rare SMBH binary hosts.
We plot the host galaxy stellar mass versus the specific star-formation rate (sSFR) of the dual AGN merger remnants at three redshift bins, overlaid on the distribution of single AGN.

In the highest redshift bin ($z=1-2.5$), DESI dual AGN mergers almost all reside in massive star-forming galaxies on the star-forming main sequence, and usually with a bright quasar ($L_{\rm bol}>10^{45.5}\sim 10^{46}\,{\rm erg/s}$).
In the intermediate redshift bin ($z=0.4-0.9$), we start seeing a significant fraction ($\sim 50\%$) of dual AGN mergers residing in green-valley galaxies with moderate sSFR ($10^{-11.5}<{\rm sSFR}<10^{-10.5}\,{\rm yr}^{-1}$) and hosting moderately luminous AGN ($L_{\rm bol}\sim 10^{44-45}\,{\rm erg/s}$), while the other half still reside in massive star-forming galaxies.
This is in line with the picture of galaxy quenching through merger-induced AGN feedback \citep[e.g.][]{DiMatteo2005}.
Notably, several dual remnants reside in the very massive end of the star-forming main sequence ($M_*>10^{11}\,M_\odot$ and ${\rm sSFR}>10^{-9.5}\,{\rm yr}^{-1}$), which are rare among the single AGN population, and could be promising SMBH binary hosts to follow up.
At low redshifts ($z=0.1-0.3$), the majority ($\sim 70\%$) of DESI dual AGN mergers are along the star-forming main sequence, making them hard to distinguish from the single AGN population.

The Roman dual remnants are more dominated by the star-forming main sequence population at all redshifts, due to the sensitivity of Roman to low-luminosity AGN in normal star-forming galaxies, although a fraction of them still reside in the green-valley region.
At higher redshifts ($z=0.5$), the fraction of dual descendants is also the highest in massive and star-forming quasar hosts, and the massive tip ($M_*\sim 10^{12}\,M_\odot$ and ${\rm sSFR}\sim 10^{-10}\,{\rm yr}^{-1}$) of the star-forming main sequence is also populated by several dual remnants.
Overall, one sweet spot to search for SMBH binary hosts evolved from dual AGN is near $z\sim 0.5$, where a significant fraction of dual remnants reside in green-valley galaxies and massive star-forming galaxies that are rare among the single AGN population.
Another promising region is the massive end of the star-forming main sequence hosting bright quasars at $z>1$, where dual remnants are over-represented compared to single AGN.

\section{Conclusions}
\label{section:conclusions}
In this work, we have constructed mock dual AGN samples tailored to current (COSMOS-Web Chandra+JWST sample and DESI dual AGN candidates) and upcoming surveys (AXIS and Roman), using the large-volume hydrodynamical simulation \astrid.
We investigated the statistical properties of these dual AGN samples, including dual fraction, separation distribution, host galaxy properties, and black hole masses.
We link the dual AGN to MBH binaries and mergers, in particular their descendants' contribution to the PTA-band GWB and to LISA-detectable GW sources.
We explored the host galaxy properties of dual AGN merger remnants to identify promising SMBH binary hosts.

We showed that when matching the selection functions of current surveys, the mock dual AGN samples from \astrid are in good agreement with observations in terms of dual fraction and host galaxy properties.
We are able to recover both the low dual fraction ($\sim 0.2\%$) in the luminous SDSS quasar sample \citep{Silverman2020ApJ...899..154S} and the higher dual fraction ($\sim 2-5\%$) in the lower-luminosity COSMOS-Web sample \citep{LiJunyao2025ApJ...986..101L} across a redshift range of $z=0-4$, using a tailored selection function for each survey.
The most recent $z\sim 3$ high dual fraction of $\sim 30\%$ reported in \cite{Perna2025A&A...696A..59P} is at $2\sigma$ above our simulation mean, when taking into account the cosmic variance and Poisson fluctuations in simulated mocks with comparable survey area.
We predict that with a high sensitivity to probe faint dual AGN down to $L_{\rm bol}\sim 10^{41}\,{\rm erg/s}$, future X-ray missions like AXIS and the infrared imaging from Roman will find $\sim 8-10\%$ AGN with a companion in the local universe ($z<0.2$).

The simulation predicts more small-separation ($<5\,{\rm kpc}$) dual AGN than currently observed in the COSMOS-Web field by \cite{LiJunyao2025ApJ...986..101L}, which could be due to observational incompleteness at small separations, or an overestimation of the AGN triggering in close galaxy pairs in the simulation.
The high resolution and large survey volume of Roman will be able to nail down this discrepancy, and constrain the secondary AGN activation fraction by providing a significant sample of small-separation dual AGN.

We showed that dual AGN are reliable tracers of SMBH mergers, with $\sim 30-70\%$ of dual AGN eventually merging by $z=0$ on a timescale of $\lesssim 1\,{\rm Gyr}$, out of which $20-60\%$ are LISA-detectable GW sources.
The exact fraction of mergers and LISA progenitors depends on the survey selection function and redshift, with Roman and Chandra+JWST-selected COSMOS-Web duals having the highest merger and LISA progenitor fraction due to their sensitivity to low-mass MBHs at small separations.
AXIS and Roman will provide significant samples of LISA progenitors: $\sim 10\%$ of low-redshift ($z<0.2$) LISA mergers will have precursors detectable as dual AGN in these two samples.

There is also overlap between dual AGN and the parent population of SMBH binaries that make up the PTA-band GWB.
We showed that $> 20\sim 40\%$ of the PTA GW background at $f=0.1 {\rm yr}^{-1}$ is contributed by MBH binaries that were in the Roman and AXIS dual AGN observation.
This fraction is lower for DESI, due to the restriction in tracer types between $z=0.5-1$, where a large fraction of SMBH binaries contribute to the PTA background.
With the large survey volume of DESI and Roman, there is also a good chance of identifying the parent population of the loudest PTA foreground sources.
Two of the six loudest sources were dual AGN in the DESI mock dual AGN sample.

We investigated the host galaxy properties of dual AGN merger remnants to identify promising SMBH binary hosts.
We found that green-valley galaxies ($10^{-11.5} < {\rm sSFR} < 10^{-10.5}\,{\rm yr}^{-1}$) with a moderately luminous AGN ($L_{\rm bol}\sim 10^{43-44}\,{\rm erg/s}$) at $z\sim 0.5$ are the most likely binary hosts evolved from dual AGN, as they are over-represented among dual remnants compared to the single AGN population.
This is in line with the picture of galaxy quenching through merger-induced AGN feedback \citep[e.g.][]{DiMatteo2005}.
Another promising region to search for SMBH binary hosts is the massive end of the star-forming main sequence ($M_*>10^{11}\,M_\odot$ and ${\rm sSFR}>10^{-9.5}\,{\rm yr}^{-1}$) that also host bright quasars ($L_{\rm bol}>10^{45.5}\,{\rm erg/s}$) at $z>1$.
Our results highlight the power of combining large-volume hydrodynamical simulations with tailored mock observations to interpret dual AGN surveys and their connection to SMBH mergers.
Future work can focus on improving the modelling of MBH dynamics and AGN triggering in galaxy mergers, to better capture the small-scale physics relevant for dual AGN formation and evolution and detailing the evolution paths from dual AGN to SMBH binaries.

\section*{Acknowledgements}
NC acknowledges support from the Schmidt Futures Fund and MPA Postdoctoral Fellowship. 
YZ and TDM acknowledge the support from the NASA FINESST grant NNH24ZDA001N. 
TDM acknowledges funding from NASA ATP 80NSSC20K0519, NSF PHY-2020295, NASA ATP NNX17AK56G, and NASA ATP 80NSSC18K101.
YN acknowledges support from the ITC Postdoctoral Fellowship.
SB acknowledges funding from NASA ATP 80NSSC22K1897.
TDM acknowledges support from NASA Theory grant 80NSSC22K072  and NASA ATP 80NSSC18K101.
\astrid~was run on the Frontera facility at the Texas Advanced Computing Center. This material is based upon work supported by the National Aeronautics and Space Administration under Grant No. 22-LPS22-0025.

This research used data obtained with the Dark Energy Spectroscopic Instrument (DESI). DESI construction and operations is managed by the Lawrence Berkeley National Laboratory. This material is based upon work supported by the U.S. Department of Energy, Office of Science, Office of High-Energy Physics, under Contract No. DE–AC02–05CH11231, and by the National Energy Research Scientific Computing Center, a DOE Office of Science User Facility under the same contract. Additional support for DESI was provided by the U.S. National Science Foundation (NSF), Division of Astronomical Sciences under Contract No. AST-0950945 to the NSF’s National Optical-Infrared Astronomy Research Laboratory; the Science and Technology Facilities Council of the United Kingdom; the Gordon and Betty Moore Foundation; the Heising-Simons Foundation; the French Alternative Energies and Atomic Energy Commission (CEA); the National Council of Humanities, Science and Technology of Mexico (CONAHCYT); the Ministry of Science and Innovation of Spain (MICINN), and by the DESI Member Institutions: www.desi.lbl.gov/collaborating-institutions. The DESI collaboration is honored to be permitted to conduct scientific research on I’oligam Du’ag (Kitt Peak), a mountain with particular significance to the Tohono O’odham Nation. Any opinions, findings, and conclusions or recommendations expressed in this material are those of the author(s) and do not necessarily reflect the views of the U.S. National Science Foundation, the U.S. Department of Energy, or any of the listed funding agencies.

\section*{Data Availability}
The code to reproduce the simulation is available at \url{https://github.com/MP-Gadget/MP-Gadget}, and continues to be developed.
The \astrid snapshots are available at \url{https://astrid-portal.psc.edu/}.
The merger catalog is available upon request.

\bibliography{main}{}
\bibliographystyle{aasjournal}

\end{document}